\documentclass{article}

\usepackage{iclr2021_conference,times}

\iclrfinalcopy
\usepackage{hyperref}
\usepackage{url}

\usepackage[utf8]{inputenc} 
\usepackage[T1]{fontenc}    
\usepackage{hyperref}       
\usepackage{url}            
\usepackage{booktabs}       
\usepackage{amsfonts}       
\usepackage{nicefrac}       
\usepackage{microtype}      
\usepackage[english]{babel}
\usepackage{natbib}

\usepackage{amsmath,amsfonts,bm}

\def\eqref#1{equation~\ref{#1}}

\def\1{\bm{1}}

\DeclareMathAlphabet{\mathsfit}{\encodingdefault}{\sfdefault}{m}{sl}
\SetMathAlphabet{\mathsfit}{bold}{\encodingdefault}{\sfdefault}{bx}{n}

\newcommand{\E}{\mathbb{E}}

\usepackage{multirow}
\usepackage{booktabs,amsfonts,dcolumn}
\usepackage{array}
\usepackage{pifont}
\usepackage{caption}
\usepackage{xspace} 
\usepackage{xcolor}
\usepackage{CJKutf8}
\usepackage{epsfig}
\usepackage{graphicx,subfigure}
\usepackage{savesym}
\usepackage{amsmath}
\usepackage{savesym}
\usepackage{amssymb}
\usepackage[normalem]{ulem}
\usepackage{floatrow, makecell}
\definecolor{midnightgreen}{rgb}{0.0, 0.29, 0.33}
\definecolor{orange}{RGB}{255,127,0}
\definecolor{darkmagenta}{RGB}{139, 0, 139}

\newfloatcommand{capbtabbox}{table}[][\FBwidth]
\usepackage{amsthm}
\usepackage{float}
\usepackage{wrapfig,booktabs}

\title{Approximate Nearest Neighbor Negative Contrastive Learning for Dense Text Retrieval}

\author{\small 
Lee Xiong\thanks{Lee and Chenyan contributed equally.}, Chenyan Xiong\footnotemark[1], Ye Li,  Kwok-Fung Tang, Jialin Liu, \\
 \small   \textbf{Paul Bennett, Junaid Ahmed, Arnold Overwijk}\\
  \small Microsoft\\
  \small \texttt{lexion, chenyan.xiong, yeli1, kwokfung.tang, jialliu,}\\
  \small \texttt{paul.n.bennett, jahmed,  arnold.overwijk@microsoft.com} \\
}

\begin{document}

\maketitle

\vspace{-0.3cm}\begin{abstract}\vspace{-0.1cm}
Conducting text retrieval in a dense representation space has many intriguing advantages. Yet the end-to-end learned dense retrieval (DR) often underperforms word-based sparse retrieval. In this paper, we first theoretically show the learning bottleneck of dense retrieval is due to the domination of uninformative negatives sampled locally in batch, which yield diminishing gradient norms, large stochastic gradient variances, and slow learning convergence. We then propose Approximate nearest neighbor Negative Contrastive Learning (ANCE), a learning mechanism that selects hard training negatives globally from the entire corpus, using an asynchronously updated ANN index. Our experiments demonstrate the effectiveness of ANCE on web search, question answering, and in a commercial search environment, showing ANCE dot-product retrieval nearly matches the accuracy of BERT-based cascade IR pipeline, while being 100x more efficient. \vspace{-0.1cm}
\end{abstract}

\section{Introduction}

Many language systems rely on text retrieval as their first step to find relevant information.
For example, search ranking~\citep{nogueira2019passage}, open domain question answering (OpenQA)~\citep{chen2017reading}, and fact verification~\citep{thorne2018fact} all first retrieve relevant documents for their later stage reranking, machine reading, and reasoning models.
All these later-stage models enjoy the advancements of deep learning techniques~\citep{rajpurkar2016squad, wang2018glue}, while, the first stage retrieval still mainly relies on matching discrete bag-of-words, e.g., BM25, which has become the bottleneck of many systems~\citep{nogueira2019passage, luan2020sparsedense, zhaotransxh2020}.

Dense Retrieval (DR) aims to overcome the sparse retrieval bottleneck by matching texts in a continuous representation space learned via deep neural networks~\citep{lee2019latent, karpukhin2020dense, luan2020sparsedense}.
It has many desired properties: fully learnable representation, easy integration with pretraining, and efficiency support from approximate nearest neighbor (ANN) search~\citep{johnson2019billion}. These make dense retrieval an intriguing potential choice to fundamentally overcome some intrinsic limitations of sparse retrieval, for example, vocabulary mismatch~\citep{croft2010search}.

\begin{wrapfigure}{tr}{0.4\linewidth}
  \centering
  \includegraphics[width=\columnwidth]{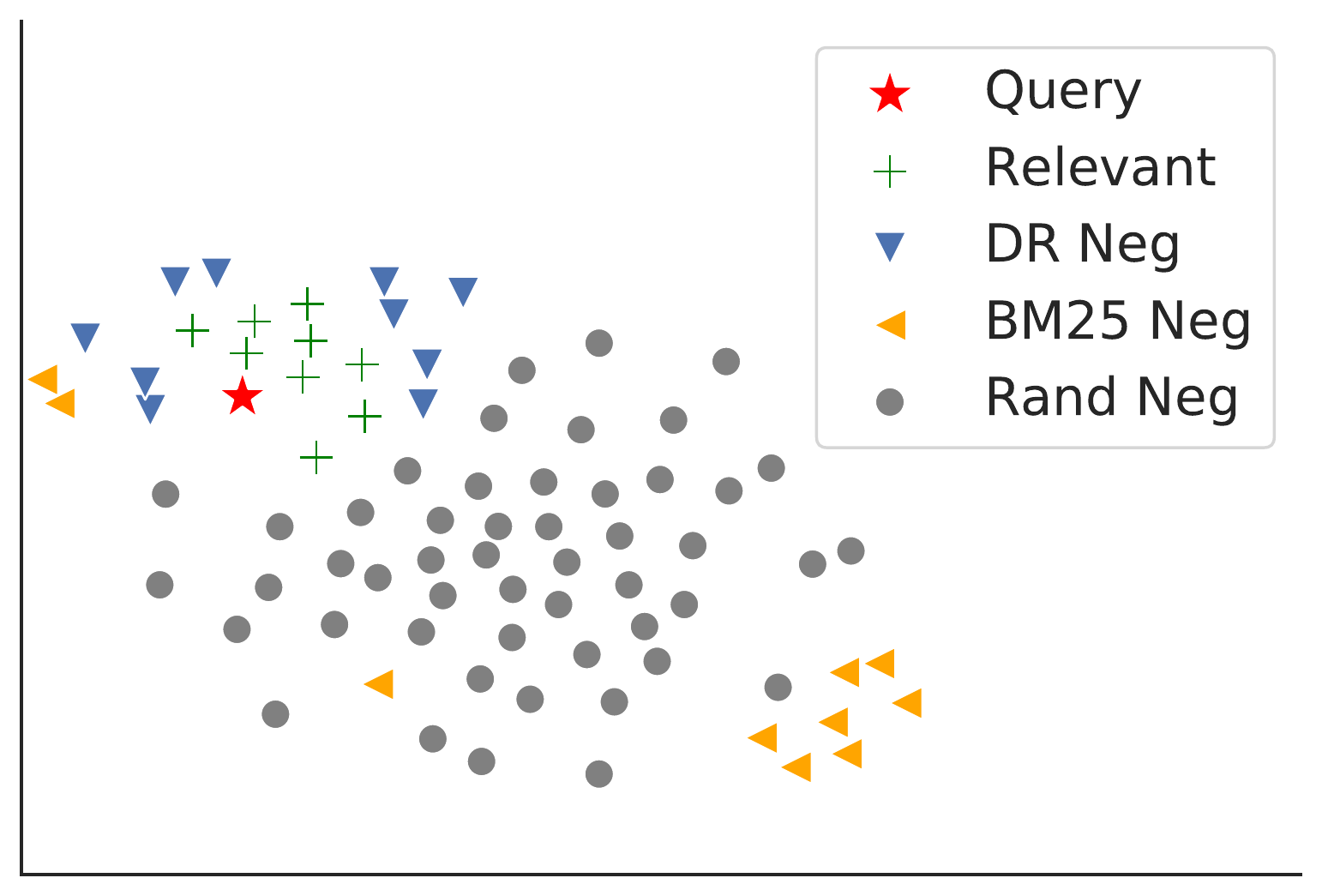}
  \caption{T-SNE~\citep{maaten2008visualizing} representations of query, relevant documents, negative training instances from BM25 (BM25 Neg) or randomly sampled (Rand Neg), and testing negatives (DR Neg) in dense retrieval.\vspace{-0.5cm}
~\label{fig.motivation} 
}

\end{wrapfigure}

A key challenge in DR is to construct proper negative instances during its representation learning~\citep{karpukhin2020dense}.
Unlike in reranking where negatives are naturally the irrelevant documents from previous retrieval stages, in first stage retrieval, DR models have to distinguish relevant documents from \textit{all irrelevant ones} in the entire corpus. As illustrated in Fig.~\ref{fig.motivation}, these \textit{global negatives} are quite different from negatives retrieved by sparse models. 

Recent research explored various ways to construct negative training instances for dense retrieval~\citep{huang2020embedding, karpukhin2020dense}., e.g., using contrastive learning~\citep{faghri2017vse++, oord2018representation, he2019momentum,  chen2020simple} to select hard negatives in current or recent mini-batches. 
However, as observed in recent research~\citep{karpukhin2020dense},
the in-batch local negatives, though effective in learning word or visual representations, are not significantly better than spare-retrieved negatives in representation learning for dense retrieval. In addition, the accuracy of dense retrieval models often underperform BM25, especially on documents~\citep{lee2019latent, gao2020complementing,luan2020sparsedense}.

In this paper, we first theoretically analyze the convergence of dense retrieval training with negative sampling.
Using the variance reduction framework~\citep{alain2015variance, katharopoulos2018not},
we show that, under conditions commonly met in dense retrieval, local in-batch negatives lead to diminishing gradient norms, resulted in high stochastic gradient variances and slow training convergence --- the local negative sampling is the bottleneck of dense retrieval's effectiveness.

Based on our analysis, we propose Approximate nearest neighbor Negative Contrastive Estimation (ANCE), a new contrastive representation learning mechanism for dense retrieval.
Instead of random or in-batch local negatives, ANCE constructs global negatives using the being-optimized DR model to retrieve from the entire corpus.
This fundamentally aligns the distribution of negative samples in training and of irrelevant documents to separate in testing.
From the variance reduction point of view, these ANCE negatives lift the upper bound of per instance gradient norm, reduce the variance of the stochastic gradient estimation, and lead to faster learning convergence.


We implement ANCE using an asynchronously updated ANN index of the corpus representation. Similar to \citet{guu2020realm}, we maintain an Inferencer that parallelly computes the document encodings with a recent checkpoint from the being optimized DR model, and refresh the ANN index used for negative sampling once it finishes, to keep up with the model training.
Our experiments demonstrate the advantage of ANCE in three text retrieval scenarios: standard web search~\citep{craswell2020overview}, OpenQA~\citep{rajpurkar2016squad, kwiatkowski2019natural}, and in a commercial search engine's retrieval system. 
We also empirically validate our theory that the gradient norms on ANCE sampled negatives are much bigger than local negatives and thus improve the convergence of dense retrieval models. Our code and trained models are available at \url{https://aka.ms/ance}.

\section{Preliminaries}
In this section, we discuss the preliminaries of dense retrieval and its representation learning.

\textbf{Task Definition:} Given a query $q$ and a corpus $C$, the first stage retrieval is to find a set of documents relevant to the query $D^+=\{d_1,...,d_i,...,d_n\}$ from $C$ ($|D^+| \ll |C|$), which then serve as input to later more complex models~\citep{croft2010search}.
Instead of using sparse term matches and inverted index, \textit{Dense Retrieval}  calculates the retrieval score $f()$ using similarities in a learned embedding space~\citep{lee2019latent, luan2020sparsedense, karpukhin2020dense}:
\begin{align}
    f(q, d) &= \text{sim}(g(q; \theta), g(d; \theta)),  \label{eq.f}
\end{align}
where $g()$ is the representation model that encodes the query or document to dense embeddings. The encoder parameter $\theta$ provides the main capacity, often fine-tuned from pretrained transformers, e.g., BERT~\citep{lee2019latent}. The similarity function (sim()) is often simply cosine or dot product, to leverage efficient ANN retrieval~\citep{johnson2019billion, guoaccelerating}.

\textbf{Learning with Negative Sampling:} The effectiveness of DR resides in learning a good representation space that maps query and relevant documents together, while separating irrelevant ones. The learning of this representation often follows standard learning to rank~\citep{liu2009learning}: Given a query $q$, a set of relevant document $D^+$ and irrelevant ones $D^-$, find the best $\theta^*$ that:
\begin{align}
    \theta^* &= \text{argmin}_{\theta} \sum_{q} \sum_{d^+\in D^+} \sum_{d^- \in D^-} l(f(q, d^+), f(q, d^-)). \label{eq.baseloss}
\end{align}
The loss $l()$ can be binary cross entropy (BCE), hinge loss, or negative log likelihood (NLL). 

A unique challenge in dense retrieval, targeting first stage retrieval, is that the irrelevant documents to separate are from the entire corpus $(D^- = C \setminus D^+)$. 
This often leads to millions of negative instances, which have to be sampled in training: 
\begin{align}
    \theta^* &= \text{argmin}_{\theta} \sum_q \sum_{d^+\in D^+} \sum_{d^- \in \hat{D}^-} l(f(q, d^+), f(q, d^-)). \label{eq.basefullloss}
\end{align}

A natural choice is to sample negatives $\hat{D}^-$ from top documents retrieved by BM25. However, they may bias the DR model to merely learn sparse retrieval and do not elevate DR models much beyond BM25~\citep{luan2020sparsedense}.
Another way is to sample negatives in local mini-batches, e.g., as in contrastive learning~\citep{oord2018representation, chen2020simple}, however, these local negatives do not significantly outperform BM25 negatives~\citep{karpukhin2020dense, luan2020sparsedense}.

\section{Analyses on The Convergence of Dense Retrieval Training}
\label{sec:theory}

In this section, we provide theoretical analyses on the convergence of representation training in dense retrieval.
We first show the connections between learning convergence and gradient norms, then the bounded gradient norms by uninformative negatives, and finally, how in-batch local negatives are ineffective under common conditions in dense retrieval.

\textbf{Convergence Rate and Gradient Norms:} Let $l(d^+, d^-)=l(f(q, d^+), f(q, d^-)$ be the loss function on the training triple $(q, d^+, d^-)$, $P_{D^-}$ the  negative sampling distribution for the given $(q, d^+)$, and $p_{d^-}$ the sampling probability of negative instance $d^-$, a stochastic gradient decent (SGD) step with importance sampling~\citep{alain2015variance} is:
\begin{align}
    \theta_{t+1} &= \theta_t - \eta \frac{1}{Np_{d^-}} \nabla_{\theta_t} l(d^+, d^-), \label{eq:wsgd}
\end{align}
with $\theta_t$ the parameter at $t$-th step, $\theta_{t+1}$ the one after, and $N$ the total number of negatives. The scaling factor $\frac{1}{Np_{d^-}}$ is to make sure Eqn.~\ref{eq:wsgd} is an unbiased estimator of the full gradient.

Then we can characterize the converge rate of this SGD step as the movement to optimal $\theta^*$. Following derivations in variance reduction~\citep{katharopoulos2018not, johnson2018training}, let $g_{d^-}= \frac{1}{Np_{d^-}} \nabla_{\theta_t} l(d^+, d^-)$ the weighted gradient, the convergence rate is:
\begin{align}
    \E\Delta^t &= ||\theta_t - \theta^*||^2 - \E_{P_{D^-}}(||\theta_{t+1} - \theta^* ||^2) \\
    &= ||\theta_t||^2 - 2\theta_t^T\theta^*  - \E_{P_{D^-}}(||\theta_{t}-\eta g_{d^-}||^2) + 2 \theta^{*T}\E_{P_{D^-}}(\theta_{t}-\eta g_{d^-}) \\
    &= -\eta^2\E_{P_{D^-}}(||g_{d^-}||^2) + 2 \eta \theta_t^T \E_{P_{D^-}}(g_{d^-}) - 2\eta \theta^{*T}\E_{P_{D^-}}(g_{d^-}) \\
    &= 2\eta \E_{P_{D^-}}(g_{d^-})^T (\theta_t - \theta^*) -\eta^2\E_{P_{D^-}}(||g_{d^-}||^2) \\
    &= 2\eta \E_{P_{D^-}}(g_{d^-})^T (\theta_t - \theta^*) -\eta^2\E_{P_{D^-}}(g_{d^-})^T \E_{P_{D^-}}(g_{d^-}) - \eta^2 \text{Tr}(\mathcal{V}_{P_{D^-}}(g_{d^-})). \label{eqn:variance}
\end{align}
This shows we can obtain better convergence rate by sampling from a distribution $P_{D^-}$ that minimizes the variance of the gradient estimator, $\E_{P_{D^-}}(||g_{d^-}||^2)$, or $\text{Tr}(\mathcal{V}_{P_{D^-}}(g_{d^-}))$ as the estimator is unbiased. There exists an optimal distribution that: 
\begin{align}
    p^*_{d^-} = \text{argmin}_{p_{d^-}} \text{Tr}(\mathcal{V}_{P_{D^-}}(g_{d^-})) \propto ||\nabla_{\theta_t} l(d^+, d^-)||_2, \label{eqn.oracle}
\end{align}
which is to sample proportionally to per instance gradient norm.
This is a well known result in importance sampling~\citep{alain2015variance, johnson2018training}. It can be proved by applying Jensen's inequality on the gradient variance and then verifying that Eqn.~\ref{eqn.oracle} achieves the minimum. We do not repeat this proof and refer to~\citet{johnson2018training} for exact derivations. 

Intuitively, an negative instance with larger gradient norm is more likely to reduce the training loss more, while those with diminishing gradients are not informative. Empirically, the correlation of gradient norm and training convergence is also observed in BERT fine-tuning~\citep{mosbach2020stability}.

\textbf{Diminishing Gradients of Uninformative Negatives:}
The oracle distribution in Eqn.~\ref{eqn.oracle} is too expensive to compute and the closed form of gradient norms can be complicated in  deep neural networks.
Nevertheless, for MLP networks, \citet{katharopoulos2018not} derives an upper bound of the per sample gradient norm:
\begin{align}
     ||\nabla_{\theta_t} l(d^+, d^-)||_2  \leq L \rho || \nabla_{\phi_L} l(d^+, d^-) ||_2,
\end{align}
where L is the number of layers, $\rho$ is composed by pre-activation weights and gradients in intermediate layers, and  $|| \nabla_{\phi_L} l(d^+, d^-) ||_2$ is the gradient w.r.t. the last layer.
Intuitively, the intermediate layers are more regulated by various normalization techniques; the main moving piece is $|| \nabla_{\phi_L} l(d^+, d^-) ||_2$~\citep{katharopoulos2018not}. 

For common learning to rank loss functions, for example, BCE loss and pairwise hinge loss, we can verified that~\citep{katharopoulos2018not}:
\begin{align}
    l(d^+, d^-) \rightarrow 0 \Rightarrow  || \nabla_{\phi_L} l(d^+, d^-) ||_2 \rightarrow 0 \Rightarrow ||\nabla_{\theta_t} l(d^+, d^-)||_2  \rightarrow 0. \label{eqn.lossbound}
\end{align}
Intuitively, negative samples with near zero loss have near zero gradients and contribute little to model convergence.
The convergence of dense retrieval model training relies on the informativeness of constructed negatives.

\textbf{Inefficacy of Local In-Batch Negatives:} We argue that the in-batch local negatives are unlikely to provide informative samples due to two common properties of text retrieval.

Let $D^{-*}$ be the set of informative negatives that are hard to distinguish from $D^{+}$, and $b$ be the batch size, we have (1) $b \ll |C|$, the batch size is far smaller than the corpus size; (2) $|D^{-*}| \ll |C|$, that only a few negatives are informative and the majority of corpus is trivially unrelated. 

Both conditions are easy to verify empirically in dense retrieval benchmarks.
The two together make the probability that a random mini-batch includes meaningful negatives $p=\frac{b |D^{-*}|}{|C|^2}$ close to zero. Selecting negatives from local training batches is unlikely to provide optimal training signals for dense retrieval.

\begin{figure}[tb]
  \centering
  \hspace*{-0cm}\includegraphics[width=0.8\columnwidth]{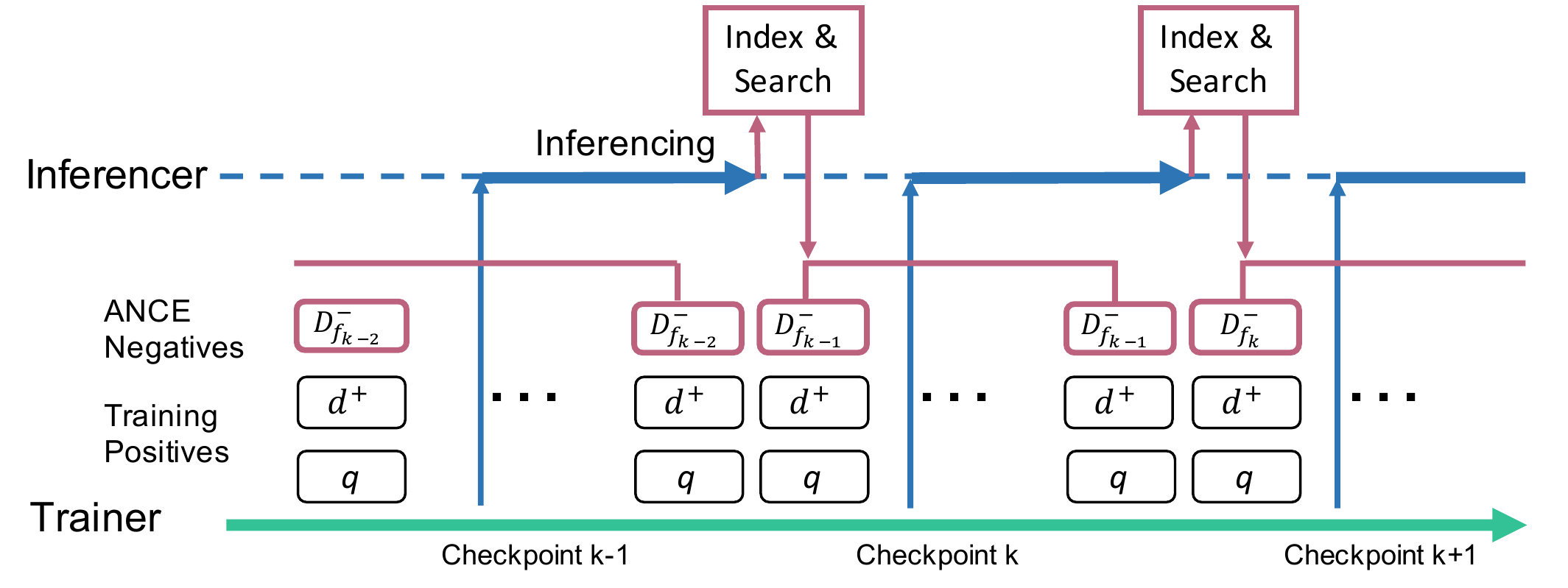}
  \caption{ANCE Asynchronous Training. The Trainer learns the representation using negatives from the ANN index. The Inferencer uses a recent checkpoint to update the representation of documents in the corpus and once finished, refreshes the ANN index with most up-to-date encodings.
  }
  \label{fig.ANCE} 
\end{figure}

\section{Approximate Nearest Neighbor Noise Contrastive Estimation}

Our analyses show the importance, if not necessity, to construct negatives \textit{globally} from the corpus. In this section, we propose \textit{A}pproximate nearest neighbor \textit{N}egative \textit{C}ontrastive \textit{E}stimation, (ANCE), which selects negatives from the entire corpus using an asynchronously updated ANN index.

\textbf{ANCE} samples negatives from the top retrieved documents via the DR model from the ANN index:
\begin{align}
  & \theta^* = \text{argmin}_{\theta} \sum_q \sum_{d^+\in D^+} \sum_{d^- \in D^-_\text{ANCE}} l(f(q, d^+), f(q, d^-)), \label{eq.fullloss}
\end{align}
with $D^-_\text{ANCE} = \text{ANN}_{f(q, d)} \setminus D^+$ and $\text{ANN}_{f(q, d)}$ the top retrieved documents by $f()$ from the ANN index.
By definition, $D^-_\text{ANCE}$ are the hardest negatives for the current DR model: $D^-_\text{ANCE} \approx D^{-*}$. In theory, these more informative negatives have higher training loss, higher upper bound on the gradient norms, and will improve training convergence.

ANCE can be used to train any dense retrieval model. For simplicity, we use a simple set up in recent research~\citep{luan2020sparsedense} with BERT Siamese/Dual Encoder (shared between $q$ and $d$), dot product similarity, and negative log likelihood (NLL) loss.

\textbf{Asynchronous Index Refresh:} 
During stochastic training, the DR model $f()$ is updated each mini-batch. Maintaining an update-to-date ANN index to select fresh ANCE negatives is challenging, as the index update requires two operations: 1) \textit{Inference}: refresh the representations of all documents in the corpus with an updated DR model;  2) \textit{Index}: rebuild the ANN index using updated representations. Although \textit{Index} is efficient~\citep{johnson2019billion}, \textit{Inference} is too expensive to compute per batch as it requires a forward pass on the entire corpus which is much bigger than the training batch. 

Thus we implement an asynchronous index refresh similar to \citet{guu2020realm}, and update the ANN index once every  $m$ batches, i.e., with checkpoint $f_k$. As illustrated in Fig.~\ref{fig.ANCE}, besides the Trainer, we run an Inferencer that takes the latest checkpoint (e.g., $f_{k}$) and recomputes the encodings of the entire corpus. In parallel, the Trainer continues its stochastic learning using $D^-_{f_{k-1}}$ from $\text{ANN}_{f_{{k-1}}}$.  Once the corpus is re-encoded, Inferencer updates the ANN index ($\text{ANN}_{f_{k}}$) and feed it to the Trainer.

In this process, the ANCE negatives ($D^-_\text{ANCE}$) are asynchronously updated to ``catch up'' with the stochastic training. 
The gap between the ANN index and the DR model optimization depends on the allocation of computing resources between Trainer and Inferencer. 
Appendix~\ref{app:lap} shows an 1:1 GPU split is sufficient to minimize the influence of this gap.

\section{Experimental Methodologies}
\label{sec:exp}
This section describes our experimental setups. More details can be found in Appendix~\ref{app:exp} and~\ref{app:hole}.

\textbf{Benchmarks:} 
The web search experiments use the TREC 2019 Deep Learning (DL) Track benchmark~\citep{craswell2020overview}, a large scale ad hoc retrieval dataset. 
We follow the official guideline and evaluate mainly in the retrieval setting, but also results when reranking top 100 BM25 candidates.


The OpenQA experiments use the Natural Questions (NQ)~\citep{kwiatkowski2019natural} and TriviaQA (TQA)~\citep{joshi2017triviaqa}, following the exact settings from \citet{karpukhin2020dense}.
The metrics are Coverage@20/100, which evaluate whether the Top-20/100 retrieved passages include the answer. We also evaluate whether ANCE's better retrieval can propagate to better answer accuracy, by running the state-of-the-art systems' readers on top of ANCE instead of DPR retrieval. The readers are RAG-Token~\citep{lewis2020retrieval} on NQ and DPR Reader on TQA, in their suggested settings.

We also study the effectiveness of ANCE in the first stage retrieval of a commercial search engine's production system. We change the training of a production-quality DR model to ANCE, and evaluate the offline gains in various corpus sizes, encoding dimensions, and exact/approximate search.

\textbf{Baselines:} In TREC DL, we include best runs in relevant categories and refer to \citet{craswell2020overview} for more baseline scores. We implement recent DR baselines that use the same BERT-Siamese, but vary in negative construction: random sampling in batch (Rand Neg), random sampling from BM25 top 100 (BM25 Neg)~\citep{lee2019latent, gao2020complementing} and the 1:1 combination of BM25 and Random negatives (BM25 + Rand Neg)~\citep{karpukhin2020dense, luan2020sparsedense}. We also compare with contrastive learning/Noise Contrastive Estimation, which uses hardest negatives in batch (NCE Neg)~\citep{gutmann2010noise, oord2018representation, chen2020simple}.
In OpenQA, we compare with DPR, BM25, and their combinations~\citep{karpukhin2020dense}.

\textbf{Implementation Details:} 
In TREC DL, recent research found MARCO passage training labels cleaner~\citep{yanidst} and BM25 negatives can help train dense retrieval~\citep{karpukhin2020dense, luan2020sparsedense}. 
Thus, we include a ``BM25 Warm Up'' setting (BM25 $\rightarrow *$), where the DR models are first trained using MARCO official BM25 Negatives. ANCE is also warmed up by BM25 negatives. All DR models in TREC DL are fine-tuned from RoBERTa base~\citep{liu2019roberta}.
In OpenQA, we warm up ANCE using the released DPR checkpoints~\citep{karpukhin2020dense}.

 To fit long documents in BERT-Siamese, ANCE uses the two settings from \citet{dai2019deeper}, FirstP which uses the first 512 tokens of the document, and MaxP, where the document is split to 512-token passages (maximum 4) and the passage level scores are max-pooled. The max-pooling operation is natively supported in ANN.
The ANN search uses the Faiss IndexFlatIP Index~\citep{johnson2019billion}.
We use 1:1 Trainer:Inference GPU allocation,  index refreshing per 10k training batches, batch size 8, and gradient accumulation step 2 on 4 GPUs. For each positive, we uniformly sample one negative from ANN top 200.
We measured ANCE efficiency using a single 32GB V100 GPU, on a cloud VM with Intel(R) Xeon(R) Platinum 8168 CPU and 650GB of RAM memory.


\begin{table*}[t]
    \centering
    \small
    \begin{tabular}{l|cc|cc|cc} \hline \hline
    & \multicolumn{2}{c|}{\textbf{MARCO Dev}} 
    & \multicolumn{2}{c|}{\textbf{TREC DL Passage}} 
    & \multicolumn{2}{c}{\textbf{TREC DL Document}}
    \\ 
    
      & \multicolumn{2}{c|}{\textbf{Passage Retrieval}} 
      & \multicolumn{2}{c|}{\textbf{NDCG@10}} 
      & \multicolumn{2}{c}{\textbf{NDCG@10}} 
    \\  \hline
    &  \textbf{MRR@10} &  \textbf{Recall@1k}
      & \textbf{Rerank} & \textbf{Retrieval}
     & \textbf{Rerank} & \textbf{Retrieval}
    \\ \hline
    
\textbf{Sparse \& Cascade IR}  &  &  &  &  & &  \\
{BM25}
& 0.240 & 0.814 & -- & 0.506 & -- & 0.519 \\
{Best DeepCT} & 0.243 & n.a. & -- & n.a. & -- & 0.554
\\
{Best TREC Trad Retrieval}
& 0.240 & n.a. & -- & 0.554 & -- & 0.549\\
{BERT Reranker}  
& 
--  & -- & \textbf{0.742} & -- & 0.646 & -- 
\\\hline

{\textbf{Dense Retrieval}}  &  &  &  &  & & \\
Rand Neg & 0.261 & 0.949 & 0.605 & 0.552 & 0.615 & 0.543 \\
NCE Neg & 0.256 & 0.943 & 0.602 & 0.539 & 0.618 & 0.542 \\
BM25 Neg & 0.299 & 0.928 & 0.664 & 0.591 & 0.626 & 0.529 \\
DPR (BM25 + Rand Neg) & 0.311 & 0.952 & 0.653 & 0.600 & 0.629 & 0.557
  \\
BM25 $\rightarrow$ Rand & 0.280 & 0.948 & 0.609 & 0.576 & 0.637 & 0.566 \\
BM25 $\rightarrow$ NCE Neg & 0.279 & 0.942 & 0.608 & 0.571 & 0.638 & 0.564 \\
BM25 $\rightarrow$ BM25 + Rand & 0.306 & 0.939 & 0.648 & 0.591 & 0.626 & 0.540 \\
\hline
{ANCE (FirstP)} & \textbf{0.330} & \textbf{0.959} & 0.677 & \textbf{0.648} & 0.641 & 0.615 \\ 
{ANCE (MaxP)} & -- & -- & -- & -- & \textbf{0.671} & \textbf{0.628} \\
\hline 
    \end{tabular}
     \caption{Results in TREC 2019 Deep Learning Track.
     Results not available are marked as ``n.a.'', not applicable are marked as ``--''. Best results in each category are marked bold. \vspace{-0.3cm}
\smallskip~\label{tab:overall}}
\end{table*}

\begin{figure}\TopFloatBoxes
   \begin{floatrow}
    \ttabbox{\resizebox{0.6\textwidth}{!}
         {
    \begin{tabular}{l|cc|cc} \hline \hline
         &   \multicolumn{2}{c|}{\textbf{Single Task}} 
         & \multicolumn{2}{c}{\textbf{Multi Task}}  \\ \hline
         
         & \multicolumn{1}{c}{\textbf{NQ}} 
         & \multicolumn{1}{c|}{\textbf{TQA}} 
         & \multicolumn{1}{c}{\textbf{NQ}} 
         & \multicolumn{1}{c}{\textbf{TQA}}  \\ \hline
         \textbf{Retriever} 
        &  {Top-20/100} 
        &  {Top-20/100}
        &  {Top-20/100}
        &  {Top-20/100} \\ \hline
        BM25 & 59.1/73.7 & 66.9/76.7 & --/-- & --/-- 
        \\ 
        DPR & 78.4/85.4 & 79.4/85.0 & 79.4/86.0 & 78.8/84.7 
        \\
        BM25+DPR & 76.6/83.8 & 79.8/84.5 &  78.0/83.9 & 79.9/84.4
        \\ \hline
        ANCE & \textbf{81.9}/\textbf{87.5} & \textbf{80.3}/\textbf{85.3} & \textbf{82.1}/\textbf{87.9} & \textbf{80.3}/\textbf{85.2} \\
        \hline 
    \end{tabular}
}}
         {  
         \caption{Retrieval results (Answer Coverage at Top-20/100) on 
         Natural Questions (NQ) and Trivial QA (TQA) in the setting from ~\citet{karpukhin2020dense}. \vspace{-0.35cm}
          \label{tab:openqa}}}
        \ttabbox{\resizebox{0.35\textwidth}{!}
     {\begin{tabular}{rrr|r} \hline \hline
\textbf{Corpus Size} & \textbf{Dim} & \textbf{Search} &
\textbf{Gain}  
\\ \hline	
250 Million & 768 & KNN & +18.4\% \\
8 Billion & 64 & KNN & +14.2\% \\
8 Billion & 64 & ANN & +15.5\% \\ \hline 
\end{tabular}}}
      {\caption{Relative gains in the first stage retrieval of a commercial search engine. The gains are from changing the training of a production DR model to ANCE. \vspace{-0.123cm}
      \label{tab.prod}
          }
      }
   
   \end{floatrow}
\end{figure}

\section{Evaluation Results}

In this section, we first evaluate the effectiveness and efficiency of ANCE. Then we empirically study the convergence of ANCE training following our theoretical analyses.

\subsection{Effectiveness and Efficiency}
\label{sec:efficiency}

The results on TREC 2019 DL benchmark are in Table~\ref{tab:overall}. ANCE significantly outperforms all sparse retrieval, including DeepCT, which uses BERT to learn term weights~\citep{dai2019transformer}. 
Among all different negative construction mechanisms, ANCE is the only one that elevates BERT-Siamese to robustly exceed the sparse methods in document retrieval.
It also outperforms DPR in passage retrieval in OpenQA (Table~\ref{tab:openqa}).
ANCE's effectiveness is even more observed in real production (Table~\ref{tab.prod}) with about 15\% relative gains all around.
Its better retrieval does indeed lead to better answer accuracy with the same readers used in RAG~\citep{lewis2020retrieval} and DPR (Table~\ref{tab.qae2e}). 

\begin{figure}\BottomFloatBoxes
   \begin{floatrow}
   
       \ttabbox{\resizebox{0.45\textwidth}{!}
         {
    \begin{tabular}{l|cc} \hline \hline
    \small
         \textbf{Model} 
         & \multicolumn{1}{c}{\textbf{NQ}} 
         & \multicolumn{1}{c}{\textbf{TQA}}  \\ \hline
        T5-11B~\citep{roberts2020much} & 34.5 & -
        \\ 
        T5-11B + SSM~\citep{roberts2020much} & 36.6 & -
        \\
        REALM~\citep{guu2020realm} & 40.4 & -
        \\ \hline
        DPR~\citep{karpukhin2020dense} & 41.5 & 56.8
        \\
        DPR + BM25~\citep{karpukhin2020dense} & 39.0 & 57.0
        \\ 
        RAG-Token~\citep{lewis2020retrieval} & 44.1 & 55.2
        \\
        RAG-Sequence~\citep{lewis2020retrieval} & 44.5 & 56.1
        \\ \hline
        ANCE + Reader  & \textbf{46.0} &  \textbf{57.5}
        \\
        \hline 
    \end{tabular}
}}
         {  
           \caption{OpenQA Test Scores in Single Task Setting. ANCE+Reader switches the retrieve of a system from DPR to ANCE and keeps the same reading model, which is RAG-Token on Natural Questions (NQ) and DPR Reader on Trivia QA (TQA).  
      \label{tab.qae2e}\vspace{-0.2cm}
      } 
      }
    
    \ttabbox{\resizebox{0.5\textwidth}{!}
         {
\begin{tabular}{l|r|r} \hline \hline
\textbf{Operation} & \textbf{Offline}  & \textbf{Online } \\ \hline
    {BM25 Index Build} & 3h & --  \\
    {BM25 Retrieval} & -- & 37ms  \\
    {BERT Rerank} &-- & 1.15s  \\
    Sparse IR Total (BM25 + BERT)  &-- & \textbf{1.42s} \\ 
    \hline
    
    {\textbf{ANCE Inference}} & \\
    {Encoding of Corpus/Per doc} & 10h/4.5ms & -- \\
    Query Encoding & --& 2.6ms \\
    ANN Retrieval (batched q) &-- & {9ms}  \\
    Dense Retrieval Total &-- & \textbf{11.6ms} 
    \\\hline
    {\textbf{ANCE Training}} &  \\
    {Encoding of Corpus/Per doc} & 10h/4.5ms & -- \\
    {ANN Index Build} & 10s & -- \\
    {Neg Construction Per Batch} & 72ms  & -- \\
    {Back Propagation Per Batch} & 19ms & -- \\
    \hline
    \end{tabular} \vspace{-1cm}

}
        \caption{Efficiency of ANCE Search and Training.\label{tab.efficiency}\vspace{-0.3cm}
        }
        }

   \end{floatrow}
\end{figure}
\begin{figure}[t]
    \centering
        \subfigure[ANCE FirstP (100\%) \label{fig:a1}]{\includegraphics[scale=0.27]{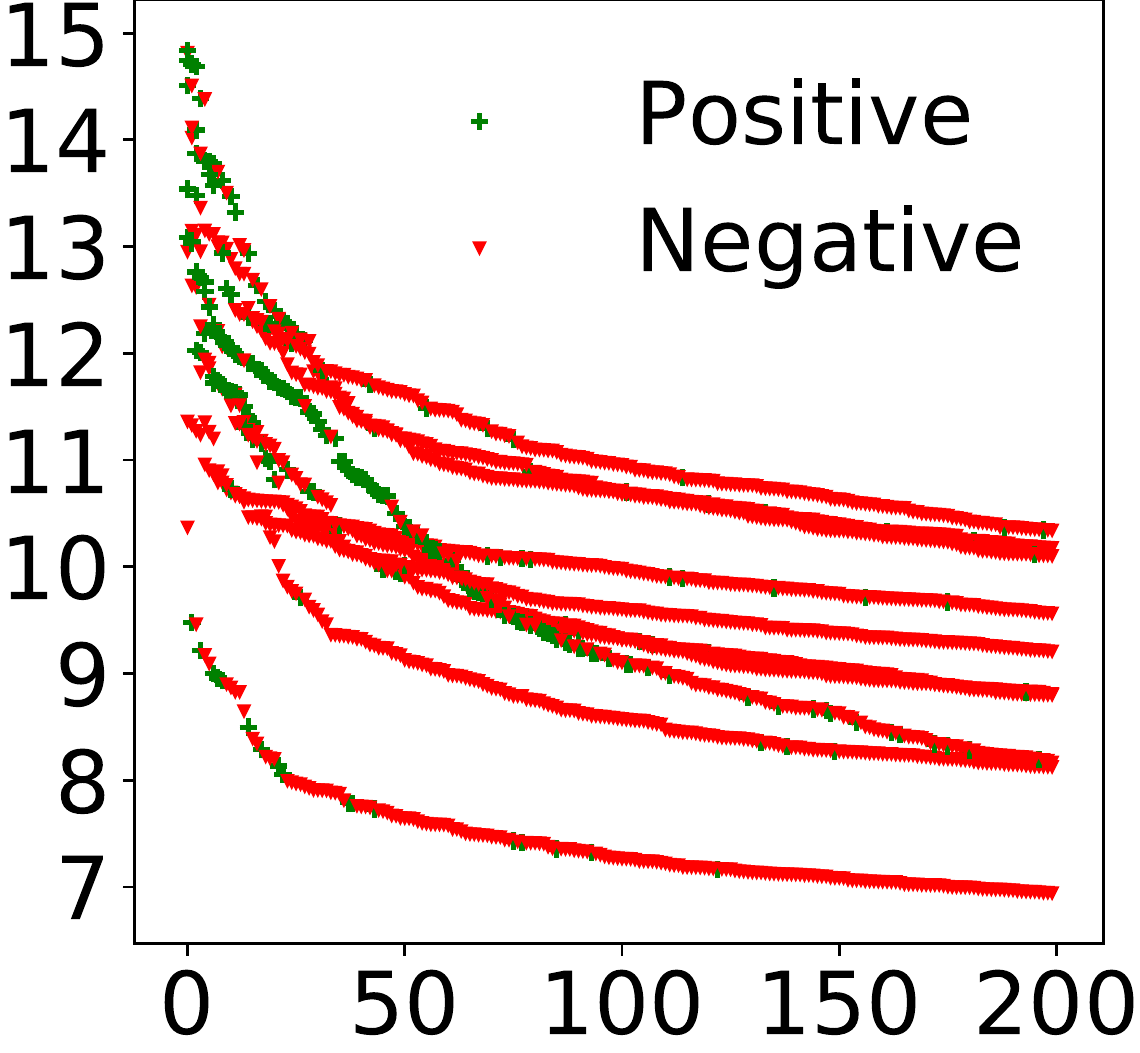}} \hspace{1mm}
        \subfigure[NCE Neg (0\%)\label{fig:b1}
        ]{\includegraphics[scale=0.27]{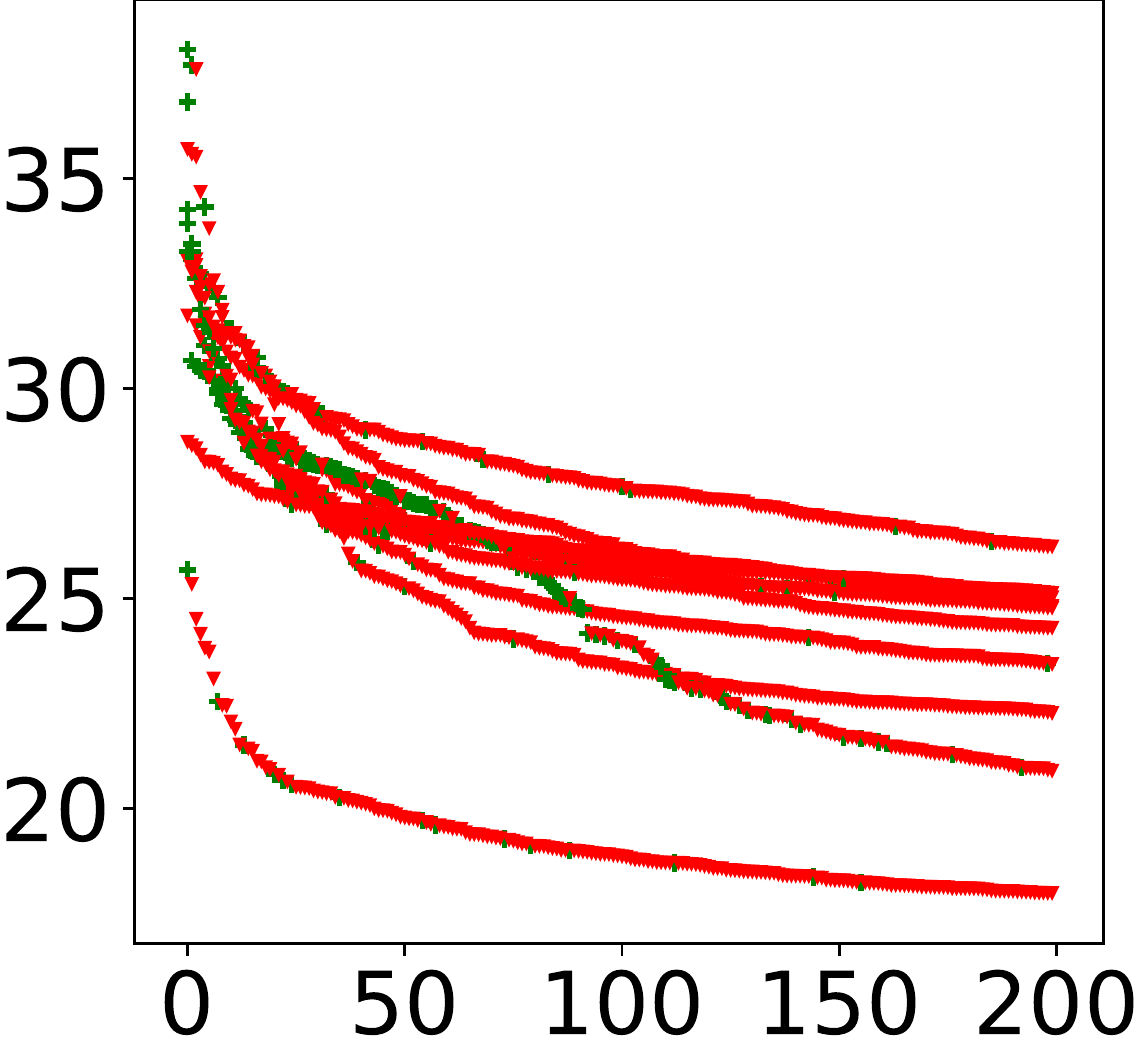}} \hspace{1mm}
        \subfigure[Rand Neg (0\%) \label{fig:c1}
        ]{\includegraphics[scale=0.27]{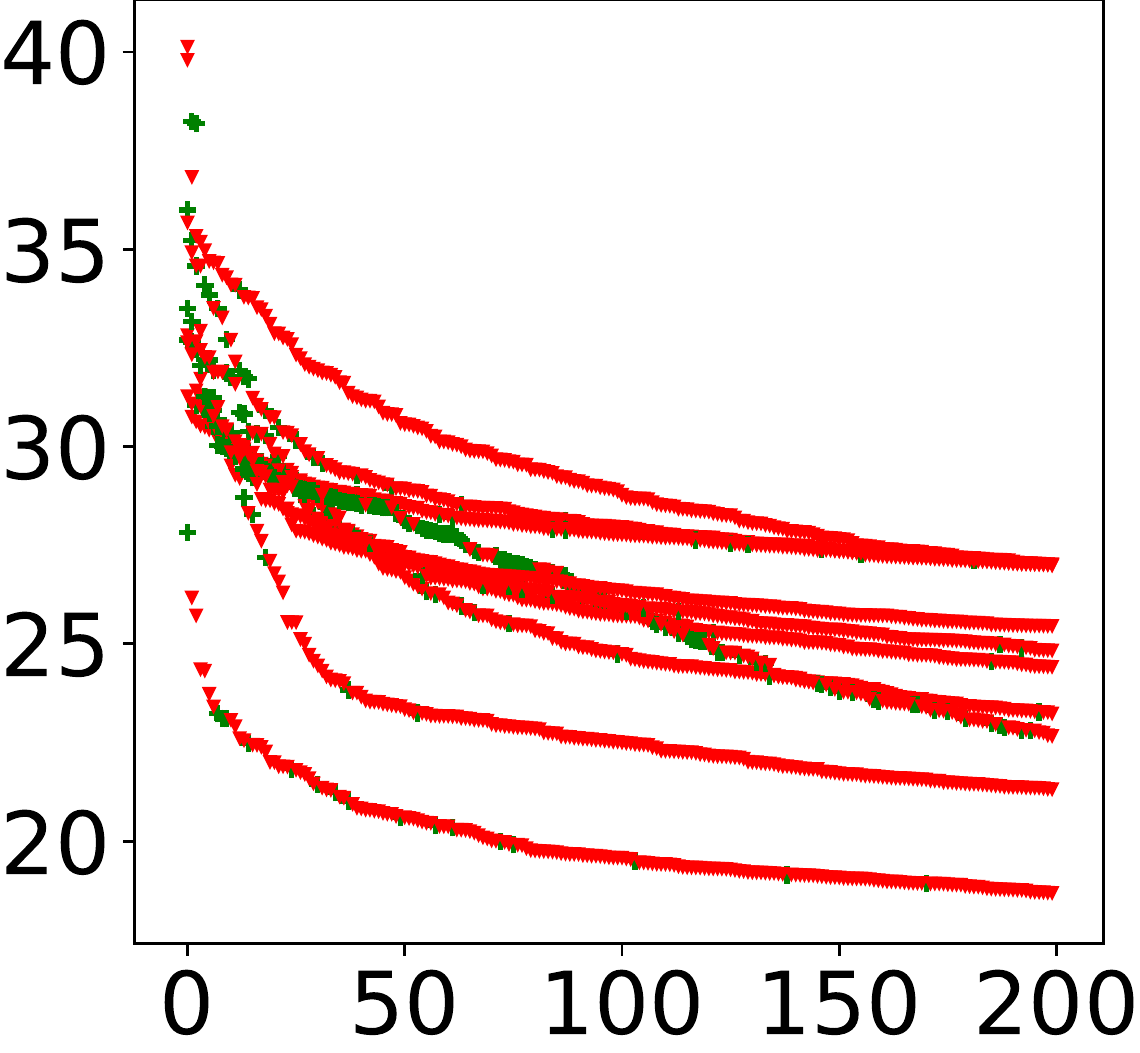}} \hspace{1mm}
        \subfigure[BM25+Rand (7\%)\label{fig:d1}      ]{\includegraphics[scale=0.27]{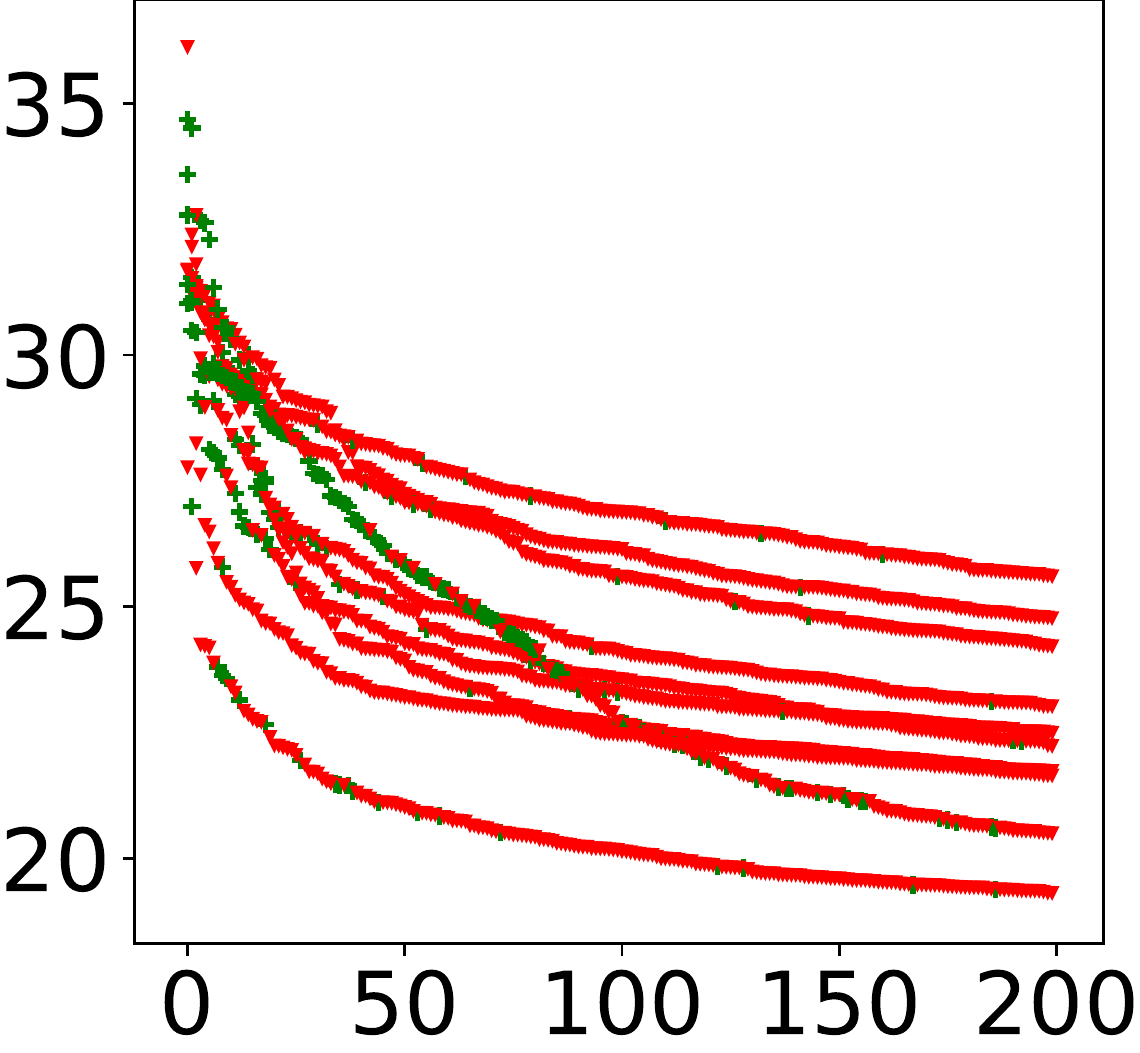}} \vspace{-0.6cm}
        \caption{ 
        The top DR scores for 10 random TREC DL testing queries. The x-axes are their ranking order. The y-axes are their retrieval scores minus corpus average.
        All models are warmed up by BM25 Neg. 
        The percentages are the overlaps between the testing and training negatives near convergence.
        \label{fig:dotprod}
        }
\end{figure} 

\begin{figure}[t]
    \centering
        \subfigure[Training Loss\label{fig:a2}      ]{\includegraphics[scale=0.27]{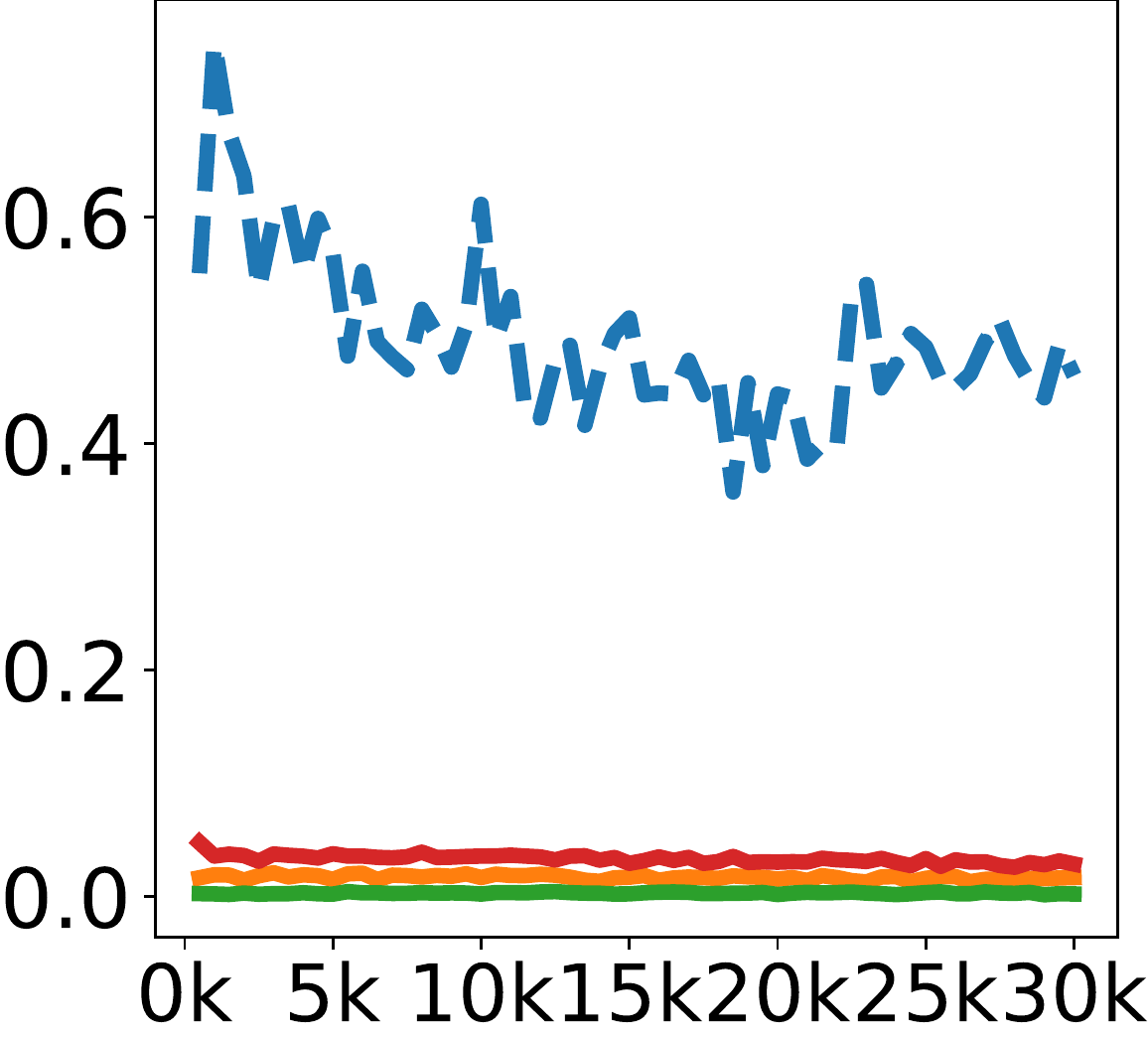}} \hspace{1mm}
        \subfigure[Grad Norm (Bottom) \label{fig:b2}]{\includegraphics[scale=0.27]{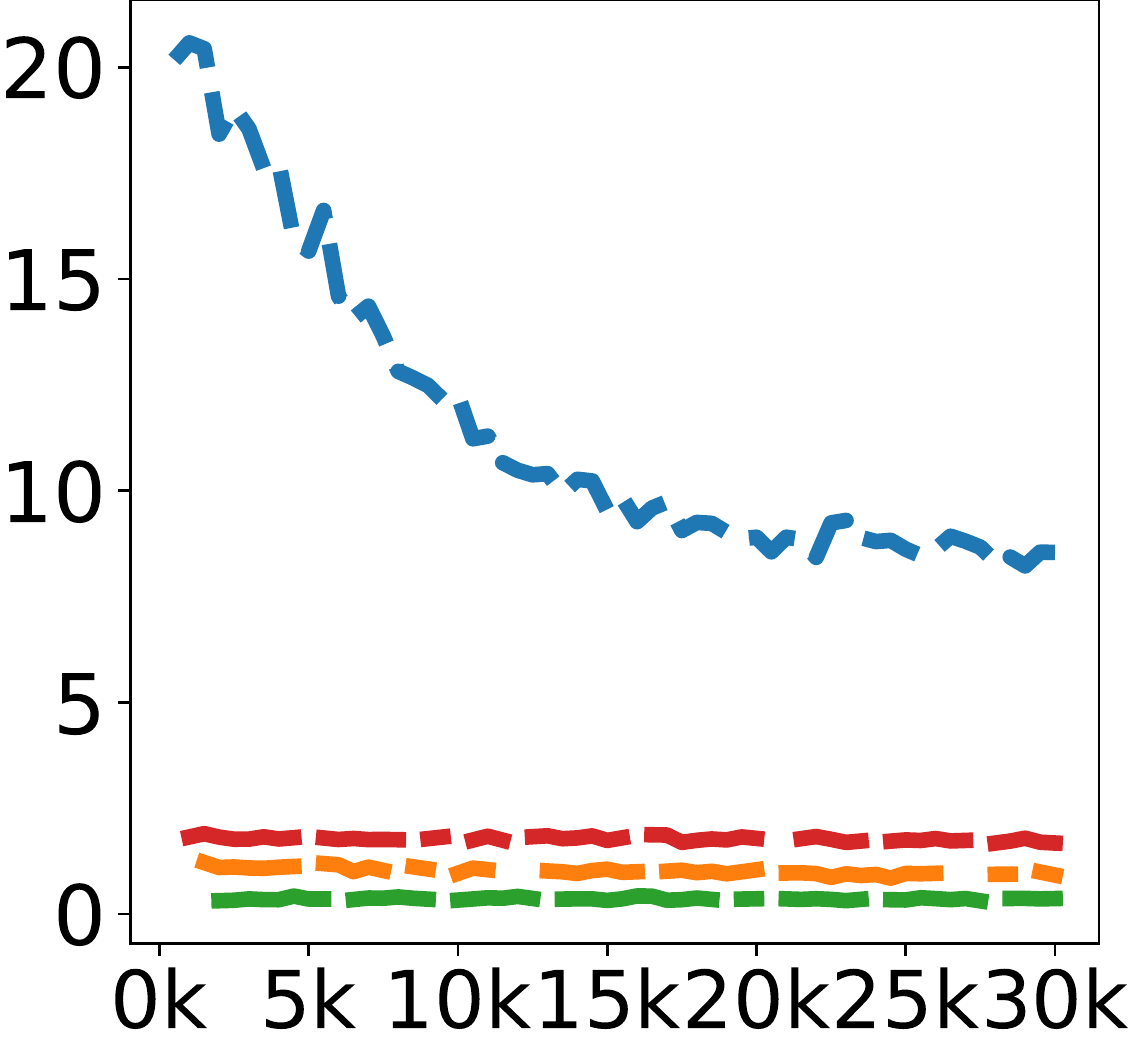}} 
        \hspace{1mm}
        \subfigure[Grad Norm (Middle)\label{fig:c2}
        ]{\includegraphics[scale=0.27]{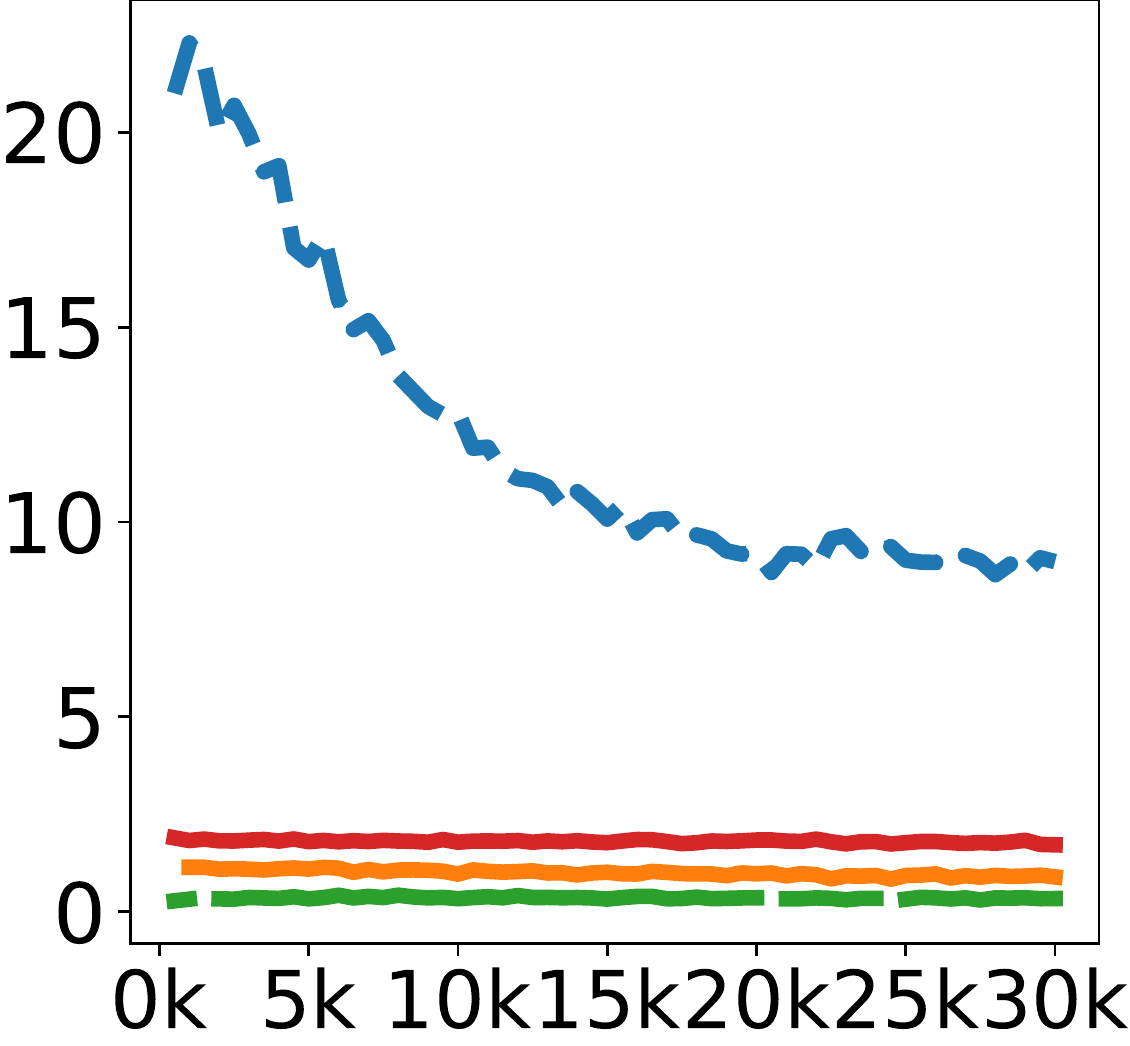}} \hspace{1mm}
        \subfigure[Grad Norm (Top)\label{fig:d2}
        ]{\includegraphics[scale=0.27]{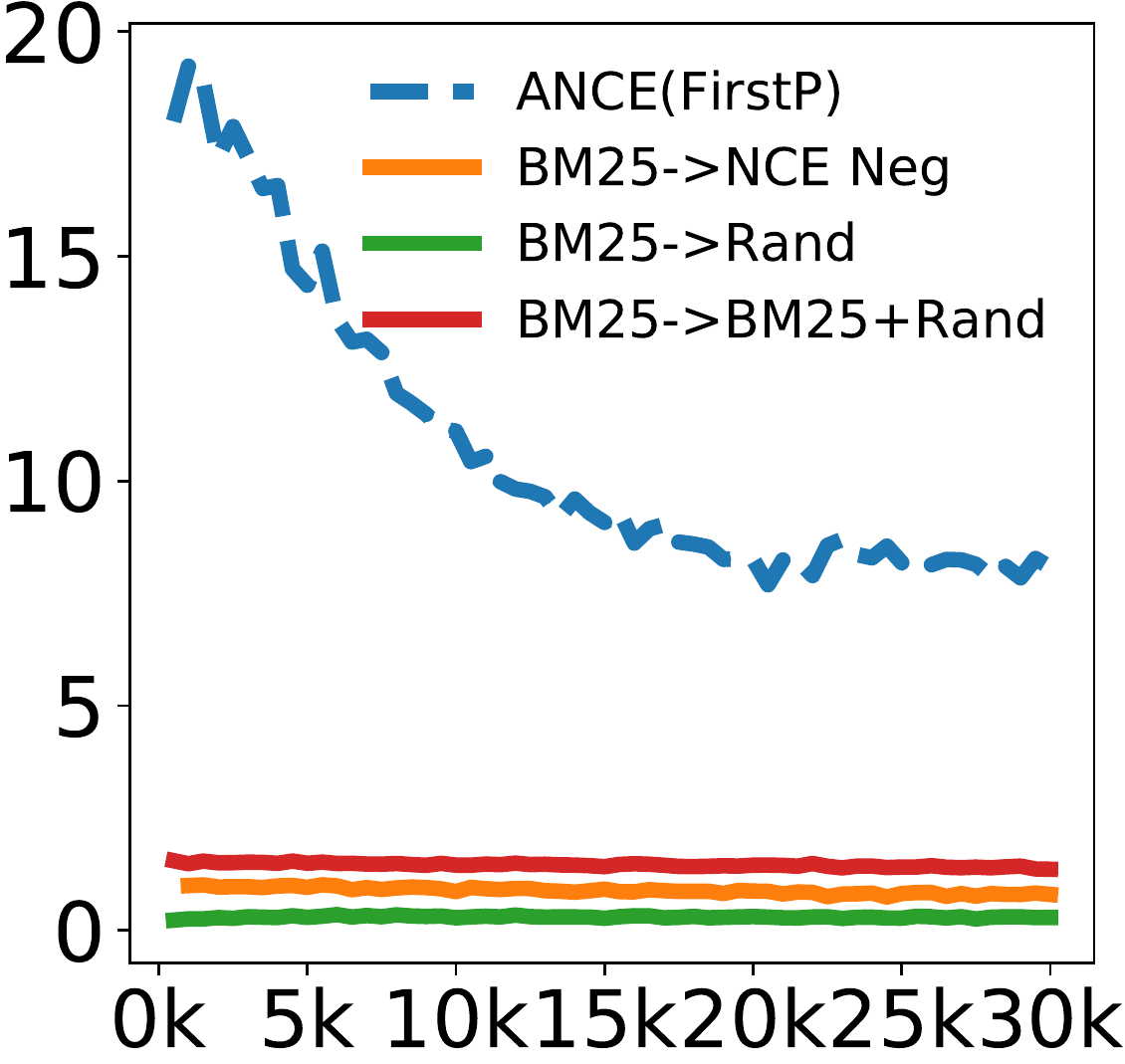}} 
        \vspace{-0.6cm}
        \caption{The loss and gradient norms during DR training (after BM25 warm up). The gradient norms are on the bottom (1-4), middle (5-8), and top (9-12) BERT layers. 
        The x-axes are training steps. 
        \label{fig:gradnorm}
        }
\end{figure} 

Among all DR models, ANCE has the smallest gap between its retrieval and reranking accuracy, showing the importance of global negatives in training retrieval models. 
ANCE retrieval nearly matches the accuracy of the cascade IR with interaction-based BERT Reranker.
This overthrows a previously-held belief that modeling term-level interactions is necessary in search~\citep{xiong2017knrm, qiao2019understanding}. 
\textit{With ANCE, we can learn a representation space that effectively captures the finesse of search relevance.}

Table~\ref{tab.efficiency} measures the efficiency ANCE (FirstP) in TREC DL document retrieval. The online latency is on one query and 100 retrieved documents.
DR with standard batching provides a \textit{100x speed up compared to BERT Rerank}, a natural benefit from the Siamese network and pre-computable document encoding. 
In ANCE training, the bulk of computing is to update the encodings of the training corpus using new checkpoints.
Assuming the model used to sample negatives and to be learned is the same, this is inevitable but can be mitigated by asynchronous index refresh.

\subsection{Empirical Analyses on Training Convergence}
\label{sec:Analyses}


We first show the long tail distribution of search relevance in dense retrieval. As plotted in Fig.~\ref{fig:dotprod}, there are a few instances per query with significant higher retrieval scores, while the majority form a long tail. 
In retrieval/ranking, the key challenge is to distinguish the relevant ones among those highest scored ones; the rest is trivially irrelevant. We also empirically measure the probability of local in-batch negatives including informative negatives ($D^{-*}$), by their overlap with top 100 highest scored negatives. 
This probability, either using NCE Neg or Rand Neg, is \textit{zero}, the same as our theory assumes. 
In comparison, the overlap between BM25 Neg with top DR retrieved negatives is 15\%, while  that of ANCE negatives starts at 63\% and converges to 100\% by design.

Then we empirically validate our theory that local negatives lead to lower loss, bounded gradient norm, and thus slow convergence.
The training loss and pre-clip gradient norms during DR training are plotted in Fig.~\ref{fig:gradnorm}. As expected, the uninformative local negatives are trivial to separate, yielding near-zero training loss, while ANCE global negatives are much harder and maintain a high training loss. 
The same with our theoretical assumption, the gradient norms of local negatives are indeed restricted closely to zero. In comparison, the gradient norms on ANCE global negatives are bigger by orders of magnitude. This confirms ANCE better approximates the oracle importance sampling distribution $ p^*_{d^-} \propto ||\nabla_{\theta_t} l(d^+, d^-)||_2$ and improves learning convergence.

\subsection{Discussions}

We use BERT-Siamese and NLL loss to be consistent with recent research. We have experimented with cosine similarity and BCE/hinge loss, where we observe even smaller gradient norms on local negatives. But the retrieval accuracy is not much better.
We include additional experiments in Appendix. \ref{app:hole} discusses the surprisingly small overlap (<25\%) between dense retrieval results and sparse retrieval results. DR is a fundamentally different approach and more studies are required to understand its behavior. 
\ref{app:lap} and~\ref{app:ablation} study the asynchronous gaps and hyperparameters. 
\ref{app:case} includes case studies that the irrelevant documents from ANCE are often still ``semantically related'' and very different from those made by sparse retrieval.

\section{Related Work}

In early research on neural information retrieval (Neu-IR)~\citep{mitra2018introduction}, a common belief was that the interaction models, those that specifically handle term level matches, are more effective though more expensive~\citep{guo2016deep, xiong2017knrm, nogueira2019passage}. 
Many techniques are developed to reduce their cost, for example, distillation~\citep{gao2020understanding} and caching~\citep{humeau2020poly, khattab2020colbert, macavaney2020efficient}. 
ANCE shows that a properly trained representation-based BERT-Siamese is in fact as effective as the interaction-based BERT ranker. This finding will motivate many new research explorations in Neu-IR.

Deep learning has been used to improve various components of sparse retrieval, for example,  term weighting~\citep{dai2019deeper}, query expansion~\citep{zheng2020bert}, and document expansion~\citep{nogueira2019document}. 
Dense Retrieval chooses a different path and conducts retrieval purely in the embedding space via ANN search~\citep{ lee2019latent,  chang2020pre, karpukhin2020dense, luan2020sparsedense}. This work demonstrates that a simple dense retrieval system can achieve SOTA accuracy, while also behaves dramatically different from classic retrieval. The recent advancement in dense retrieval may raise a new generation of search systems.

Recent research in contrastive representation learning also shows the benefits of sampling negatives from a larger candidate pool.
In computer vision, \citet{he2019momentum} decouple the negative sampling pool size with training batch size, by maintaining a negative candidate pool of recent batches and updating their representation with momentum. This enlarged negative pool significantly improves unsupervised visual representation learning~\citep{chen2020improved}.
A parallel work~\citep{xiong2020answering} improves DPR by sampling negatives from a memory bank~\citep{wu2018unsupervised} --- in which the representations of negative candidates are frozen so more candidates can be stored.
Instead of a bigger local pool, ANCE goes all the way along this trajectory and constructs negatives globally from the entire corpus, using an asynchronously updated ANN index.

Besides being a real world application itself, dense retrieval is also a core component in many other language systems, for example, to retrieval relevant information for grounded language models~\citep{khandelwal2019generalization, guu2020realm},  extractive/generative QA~\citep{karpukhin2020dense, lewis2020retrieval}, and fact verification~\citep{xiong2020answering}, or to find paraphrase pairs for pretraining~\citep{lewis2020pre}. There dense retrieval models are either frozen or optimized indirectly by signals from their end tasks.
ANCE is orthogonal with those lines of research and focuses on the representation learning for dense retrieval. Its better retrieval accuracy can benefit many language systems.

\section{Conclusion}


In this paper, we first provide theoretical analyses on the convergence of representation learning in dense retrieval. 
We show that under common conditions in text retrieval, the local negatives used in DR training are uninformative, yield low gradient norms, and contribute little to the learning convergence.
We then propose ANCE to eliminate this bottleneck by constructing training negatives globally from the entire corpus.
Our experiments demonstrate the advantage of ANCE in web search, OpenQA, and the production system of a commercial search engine.
Our studies empirically validate our theory that ANCE negatives have much bigger gradient norms, reduce the stochastic gradient variance, and improve training convergence.

\newpage

\small
\bibliography{ms.bib}
\bibliographystyle{iclr2021_conference}

\newpage
\appendix
\section{Appendix}

\subsection{More Experimental Details}
\label{app:exp}

\textbf{More Details on TREC DL Benchmarks:} 
There are two tasks in the TREC DL 2019 Track: document retrieval and passage retrieval. The training and development sets are from MS MARCO, which includes passage level relevance labels for one million Bing queries~\citep{bajaj2016ms}. The document corpus was post-constructed by back-filling the body texts of the passage's URLs and their labels were inherited from its passages~\citep{craswell2020overview}. 
The testing sets are labeled by NIST accessors on the top 10 ranked results from past Track participants~\citep{craswell2020overview}.

TREC DL official metrics include NDCG@10 on test and MRR@10 on MARCO Passage Dev. MARCO Document Dev is noisy and the recall on the DL Track testing is less meaningful due to low label coverage on DR results.
There is a two-year gap between the construction of the passage training data and the back-filling of their full document content. Some original documents were no longer available. There is also a decent amount of content changes in those documents during the two-year gap, and many no longer contain the passages. 
This back-filling perhaps is the reason why many Track participants found the passage training data is more effective than the inherited document labels. Note that the TREC testing labels are not influenced as the annotators were provided the same document contents when judging.

All the TREC DL runs are trained using these training data. Their inference results on the testing queries of the document and the passage retrieval tasks were evaluated by NIST assessors in the standard TREC-style pooling technique~\citep{voorhees2000variations}. 
The pooling depth is set to 10, that is, the top 10 ranked results from all participated runs are evaluated, and these evaluated labels are released as the official TREC DL benchmarks for passage and document retrieval tasks.

\textbf{More Details on OpenQA Experiments:} 
All the DPR related experimental settings, baseline systems, and DPR Reader are based on their open source libarary\footnote{https://github.com/facebookresearch/DPR.}. The RAG-Token reader uses their open-source release in huggingface\footnote{https://huggingface.co/transformers/master/model\_doc/rag.html}. The RAG-Seq release in huggingface is not yet stable by the time we did our experiment, thus we choose the RAG-Token in our OpenQA experiment. RAG only releases the NQ models thus we use DPR reader on TriviaQA.
We feed top 20 passages from ANCE to RAG-Token on NQ and top 100 passages to DPR's BERT Reader, following the guideline in their open-source codes. 

\textbf{More Details on Baselines:}
The most representative sparse retrieval baselines in TREC DL include the standard BM25 (``bm25base'' or ``bm25base\_p''), Best TREC Sparse Retrieval (``bm25tuned\_rm3'' or ``bm25tuned\_prf\_p'') with tuned query expansion~\citep{lavrenko2017relevance}, and Best DeepCT (``dct\_tp\_bm25e2'', doc only), which uses BERT to estimate the term importance for BM25~\citep{dai2019context}. These three runs represent the standard sparse retrieval, best classical sparse retrieval, and the recent progress of using BERT to improve sparse retrieval.
We also include the standard cascade retrieval-and-reranking systems BERT Reranker (``bm25exp\_marcomb'' or ``p\_exp\_rm3\_bert''), which is the best run using standard BERT on top of query/doc expansion, from the groups with multiple top MARCO runs~\citep{nogueira2019passage, nogueira2019document}.

\textbf{BERT-Siamese Configurations:} We follow the network configurations in ~\citet{luan2020sparsedense} in all Dense Retrieval methods, which we found provides the most stable results. 
More specifically, we initialize the BERT-Siamese model with RoBERTa base~\citep{liu2019roberta} and add a $768\times768$ projection layer on top of the last layer's ``[CLS]'' token, followed by a layer norm.

\textbf{Implementation Details:}
The training often takes about 1-2 hours per ANCE epoch, which is whenever new ANCE negative is ready, it immediately replaces existing negatives in training, without waiting.
It converges in about 10 epochs, similar to other DR baselines. The optimization uses LAMB optimizer, learning rate 5e-6 for document and 1e-6 for passage retrieval, and linear warm-up and decay after 5000 steps. More detailed hyperparameter settings can be found in our code release.

\begin{table*}[t]
    \centering
    \small
     \resizebox{\linewidth}{!}{
    \begin{tabular}{l|ccc|ccc} \hline \hline
    & \multicolumn{3}{c|}{\textbf{TREC DL Passage}} 
    & \multicolumn{3}{c}{\textbf{TREC DL Document}}
    \\ 
    \hline
    
  \textbf{Method}
  & \textbf{Recall@1K} & \textbf{Hole@10} & \textbf{Overlap w. BM25}
       & \textbf{Recall@100} & \textbf{Hole@10} & \textbf{Overlap w. BM25}
  \\  \hline
    BM25 &0.685 & 5.9\% & 100\% &0.387 &0.2\% & 100\% \\ 
    \hline
    BM25 Neg &0.569 &25.8\% &11.9\%  &0.217 &28.1\% &17.9\%  \\
    BM25 + Rand Neg  &0.662 &20.2\% &16.4\%  &0.240 &21.4\% &21.0\%  \\ \hline
    ANCE (FirstP) &0.661 &14.8\% &17.4\%  &0.266 &13.3\% &24.4\%  \\
    ANCE (MaxP) &- &- &-  &0.286 &11.9\% &24.9\%  \\ \hline
  \end{tabular}}
  \caption{Coverage of TREC 2019 DL Track labels on Dense Retrieval methods. Overlap with BM25 is calculated on top 100 retrieved documents.
  \label{tab:hole}}
\end{table*}
\subsection{Overlap with Sparse Retrieval in TREC 2019 DL Track}
\label{app:hole}

As a nature of TREC-style pooling evaluation, only those ranked in the top 10 by the 2019 TREC participating systems were labeled. As a result, documents not in the pool and thus not labeled are all considered irrelevant, even though there may be relevant ones among them. When reusing TREC style relevance labels, it is very important to keep track of the ``hole rate'' on the evaluated systems, i.e., the fraction of the top K ranked results without TREC labels (not in the pool). A larger hole rate shows that the evaluated methods are very different from those systems that participated in the Track and contributed to the pool, thus the evaluation results are not perfect. Note that the hole rate does not necessarily reflect the accuracy of the system, only the difference of it.

In TREC 2019 Deep Learning Track, all the participating systems are based on sparse retrieval.
Dense retrieval methods often differ considerably from sparse retrievals and in general will retrieve many new documents. This is confirmed in Table~\ref{tab:hole}. All DR methods have very low overlap with the official BM25 in their top 100 retrieved documents. At most, only 25\% of documents retrieved by DR are also retrieved by BM25.  This makes the hole rate quite high and the recall metric not very informative. It also suggests that DR methods might benefit more in this year's TREC 2020 Deep Learning Track if participants are contributing DR based systems.

The MS MARCO ranking labels were not constructed based on pooling the sparse retrieval results. They were from Bing~\citep{bajaj2016ms}, which uses many signals beyond term overlap. This makes the recall metric in MS MARCO more robust as it reflects how a single model can recover a complex online system.

\subsection{Impact of Asynchronous Gap}
\label{app:lap}

\begin{figure}[t]
    \centering
        \subfigure[10k Batch; 4:4; 1e-5\label{fig:a}]{\includegraphics[scale=0.27]{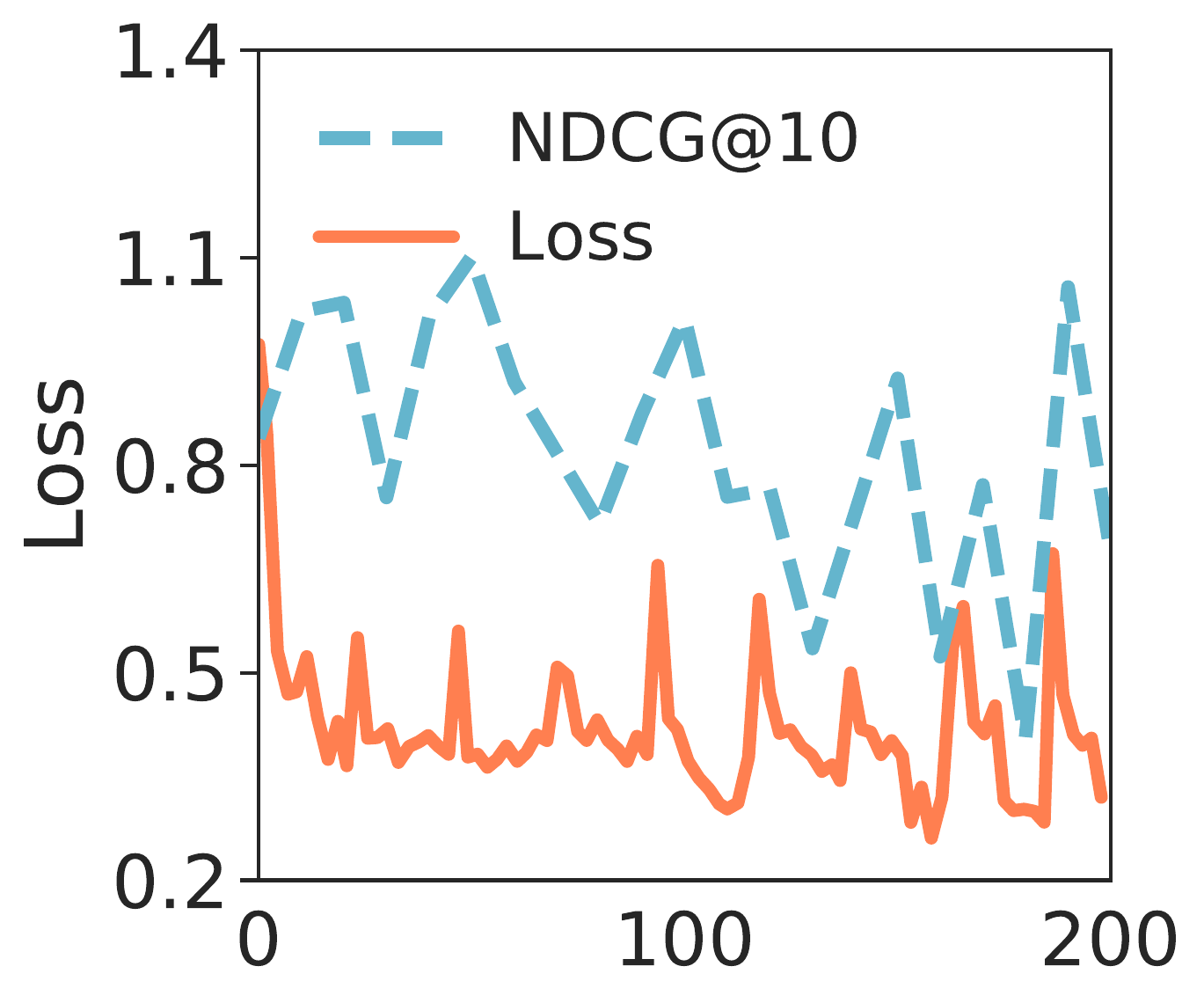}}
        \subfigure[20k Batch; 8:4; 1e-6\label{fig:b}
        ]{\includegraphics[scale=0.27]{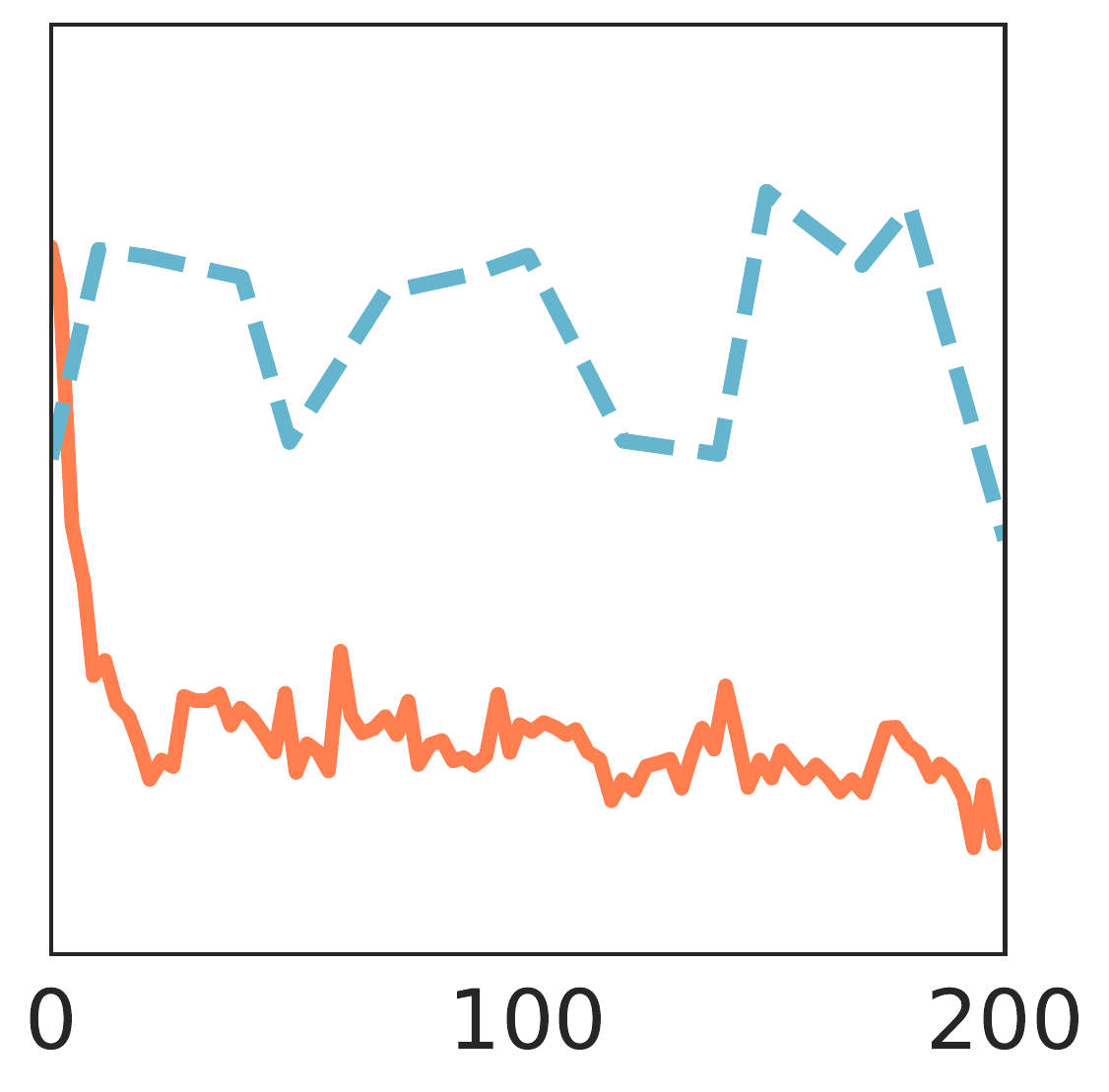}}
        \subfigure[5k Batch; 4:8; 1e-6\label{fig:c}
        ]{\includegraphics[scale=0.27]{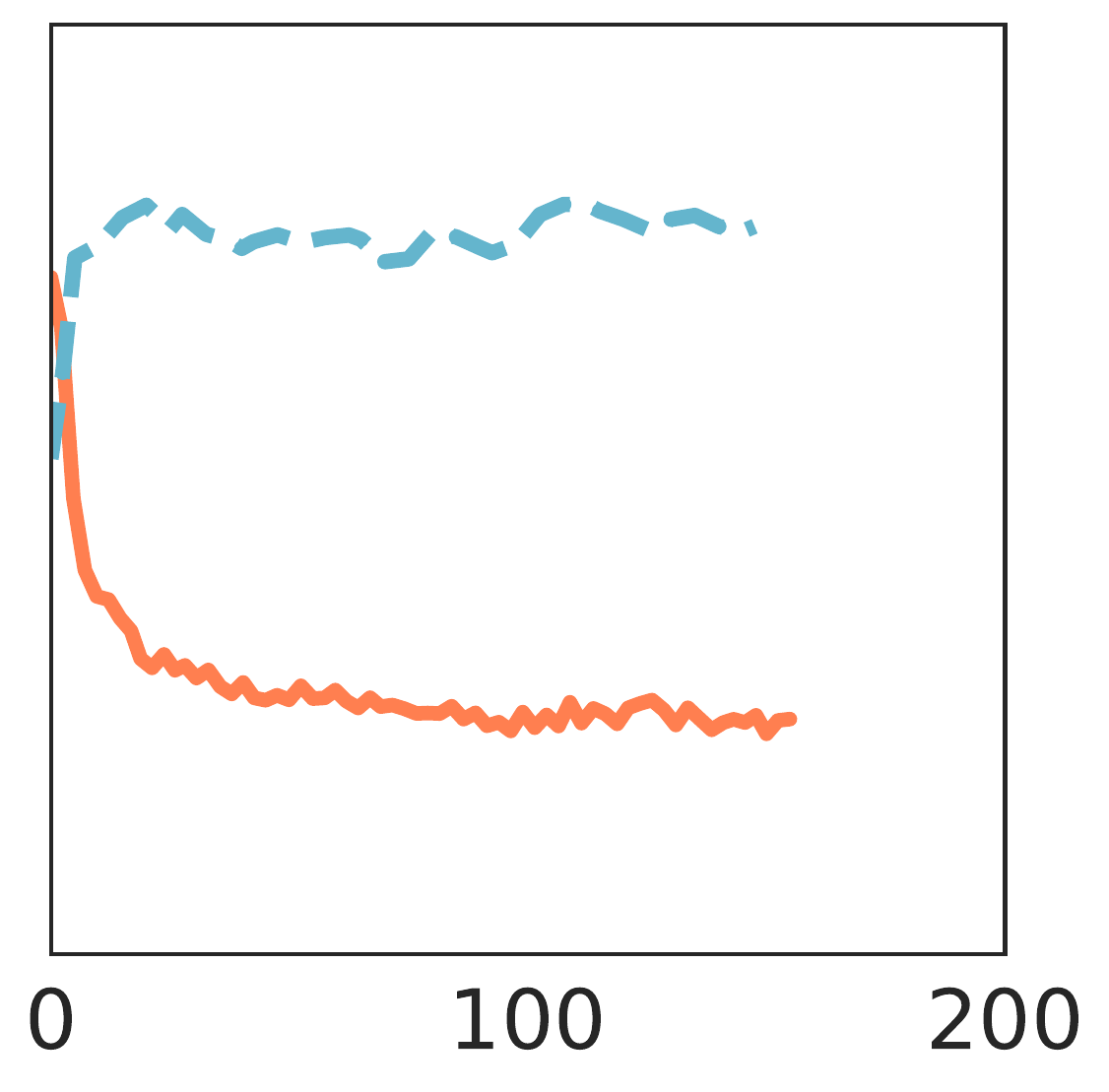}}
        \subfigure[10k Batch; 4:4; 5e-6\label{fig:d}      ]{\includegraphics[scale=0.27]{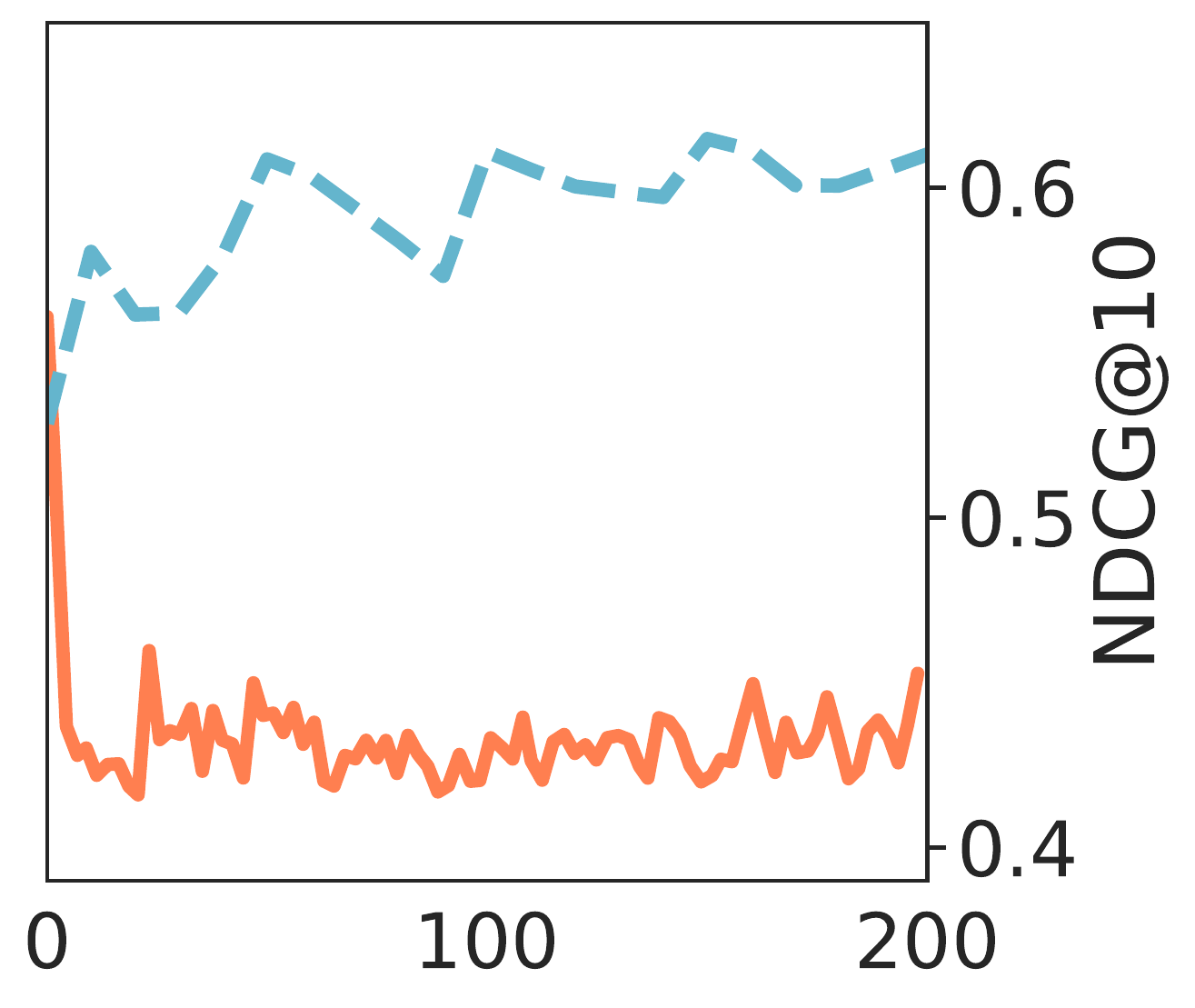}}
        \caption{
        Training loss and testing NDCG of ANCE (FirstP) on documents, with different ANN index refreshing (e.g., per 10k Batch), Trainer:Inferencer GPU allocation, and learning rate (e.g., 1e-5). X-axes is the training steps in thousands.
        \label{fig:fluct}
        }
\end{figure}

Fig.~\ref{fig:fluct} illustrates the behavior of asynchronous learning with different configurations.
A large learning rate or a low refreshing rate (Figure~\ref{fig:a} and~\ref{fig:b}) leads to fluctuations as the async gap of the ANN index may drive the representation learning to undesired local optima.
Refreshing as often as every 5k Batches yields a smooth convergence (Figure~\ref{fig:c}), but requires twice as many GPU allocated to the Inferencer.
A 1:1 GPUs allocation of Trainer and Inference with appropriate learning rates is adequate to minimize the impact of async gap.

\subsection{Hyperparameter Studies}
\label{app:ablation}

\begin{table*}[t]
    \centering
    \small
     \resizebox{\linewidth}{!}{
    \begin{tabular}{l|ccc|c|c} \hline \hline
    & \multicolumn{3}{c|}{\textbf{Hyperparameter}} 
    & \multicolumn{1}{c|}{\textbf{MARCO Dev Passage}} 
    & \multicolumn{1}{c}{\textbf{TREC DL Document}}
    \\ 
      & {\textbf{Learning rate}}
      & {\textbf{Top K Neg}}
      & {\textbf{Refresh (step)}}
      
      & {\textbf{Retrieval MRR@10}} 
      & {\textbf{Retrieval NDCG@10}} 
    \\ \hline
    
\textbf{Passage ANCE} 
& 1e-6 & 200 & 10k & \textbf{0.33} & -- \\
& 1e-6 & 500 & 10k & 0.31 & -- \\
& 2e-6 & 200 & 10k & 0.29 & -- \\
& 2e-7 & 500 & 20k & 0.303 & -- \\
& 2e-7 & 1000 & 20k & 0.302 & -- \\

\hline 
\textbf{Document ANCE} 
& 1e-5 & 100 & 10k & -- & 0.58 \\
& 1e-6 & 100 & 20k & -- & 0.59 \\
& 1e-6 & 100 & 5k & -- & 0.60 \\
& 5e-6 & 200 & 10k & -- & \textbf{0.614} \\
& 1e-6 & 200 & 10k & -- & 0.61 \\
\hline
    \end{tabular}}
     \caption{Results of several different hyperparameter configurations. ``Top K Neg'' lists the top k ANN retrieved candidates from which we sampled the ANCE negatives from.
     ~\label{tab:ablation}}
\end{table*}



We show the results of some hyperparameter configurations in Table~\ref{tab:ablation}. The cost of training with BERT makes it difficult to conduct a lot hyperparameter exploration.
Often a failed configuration leads to divergence early in training.
We barely explore other configurations due to the time-consuming nature of working with pretrained language models. Our DR model architecture is kept consistent with recent parallel work and the learning configurations in Table~\ref{tab:ablation} are about all the explorations we did. Most of the hyperparameter choices are decided solely using the training loss curve and otherwise by the loss in the MARCO Dev set. We found the training loss, validation NDCG, and testing performance align well in our (limited) hyperparameter explorations.

\begin{table}[t]
\centering
\small
\begin{tabular}{l|p{5cm}|p{5cm}}

 \hline\hline
   & \textbf{ANCE} 
    & \textbf{BM25}
    \\
     \hline

\textbf{Query:} &\multicolumn{2}{l} {qid (104861): Cost of interior concrete flooring}
\\ \hline
{Title:} &  Concrete network: Concrete Floor Cost & Pinterest: Types of Flooring \\ \hline
{DocNo:} &D293855  &D2692315  \\ \hline
{Snippet:} & For a concrete floor with a basic finish, you can expect to pay \$2 to \$12 per square foot… 
& 
Know About Hardwood Flooring And Its Types White Oak Floors Oak Flooring Laminate Flooring In Bathroom …  \\  \hline

{Ranking Position:} & 1 & 1 \\ \hline
{TREC Label:} & 3 (Very Relevant) & 0 (Irrelevant) \\ \hline
{NDCG@10:} & 0.86 & 0.15\\ \hline\hline

\textbf{Query:} &\multicolumn{2}{l} {qid (833860): What is the most popular food in Switzerland}
\\ \hline
{Title:} &  Wikipedia: Swiss cuisine & Answers.com: Most popular traditional food dishes of Mexico \\ \hline
{DocNo:} &D1927155   &D3192888   \\ \hline
{Snippet:} & Swiss cuisine bears witness to many regional influences, … Switzerland was historically a country of farmers, so traditional Swiss dishes tend not to be… 
& 
One of the most popular traditional Mexican deserts is a spongy cake … (in the related questions section) What is the most popular food dish in Switzerland?…  \\  \hline

{Ranking Position:} & 1 & 1 \\ \hline
{TREC Label:} & 3 (Very Relevant) & 0 (Irrelevant) \\ \hline
{NDCG@10:} & 0.90 & 0.14\\ \hline\hline

\textbf{Query:} &\multicolumn{2}{l} {qid (1106007): Define visceral}
\\ \hline
{Title:} &  Vocabulary.com: Visceral & Quizlet.com: A\&P EX3 autonomic 9-10 \\ \hline
{DocNo:} &D542828    &D830758    \\ \hline
{Snippet:} & When something's visceral, you feel it in your guts. A visceral feeling is intuitive — there might not be a rational explanation, but you feel that you know what's best… 
& 
Acetylcholine A neurotransmitter liberated by many peripheral nervous system neurons and some central nervous system neurons…  \\  \hline

{Ranking Position:} & 1 & 1 \\ \hline
{TREC Label:} & 3 (Very Relevant) & 0 (Irrelevant) \\ \hline
{NDCG@10:} & 0.80 & 0.14\\ \hline

\end{tabular}
\caption{Queries in the TREC 2019 DL Track Document Ranking Tasks where ANCE performs better than BM25. Snippets are  manually extracted. The documents in the first disagreed ranking position are shown, where on all examples ANCE won.
The NDCG@10 of ANCE and BM25 in the corresponding query is listed.}
\label{tab:win_case_study}
\end{table}

\begin{figure}[t]
    \centering
        \subfigure[104861: interior flooring cost. ]{\label{case_ConcreteFloor}\includegraphics[scale=0.285]{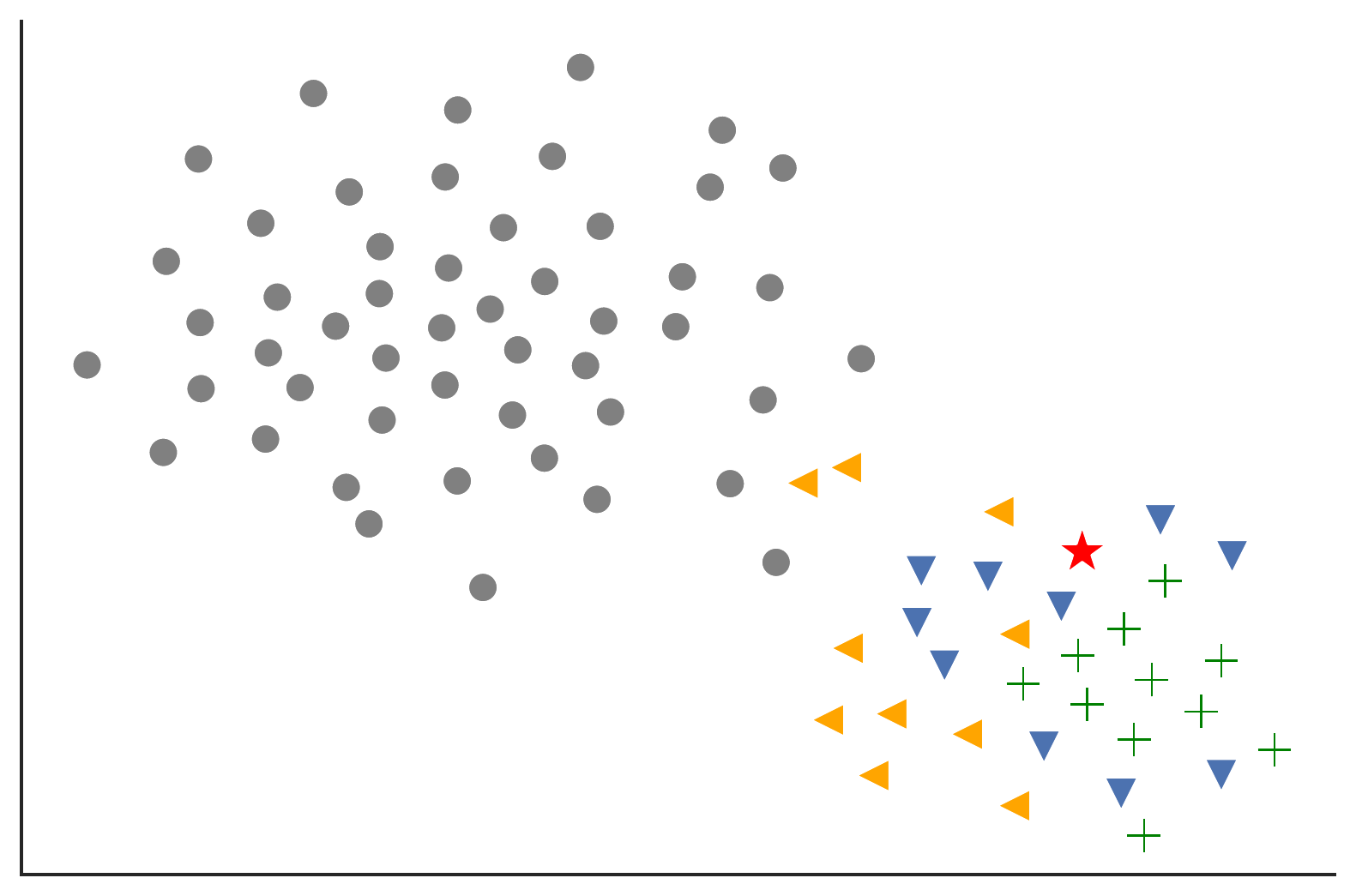}}
        \subfigure[833860: popular Swiss food]
        {\label{case_SwitzerlandFood}\includegraphics[scale=0.285]{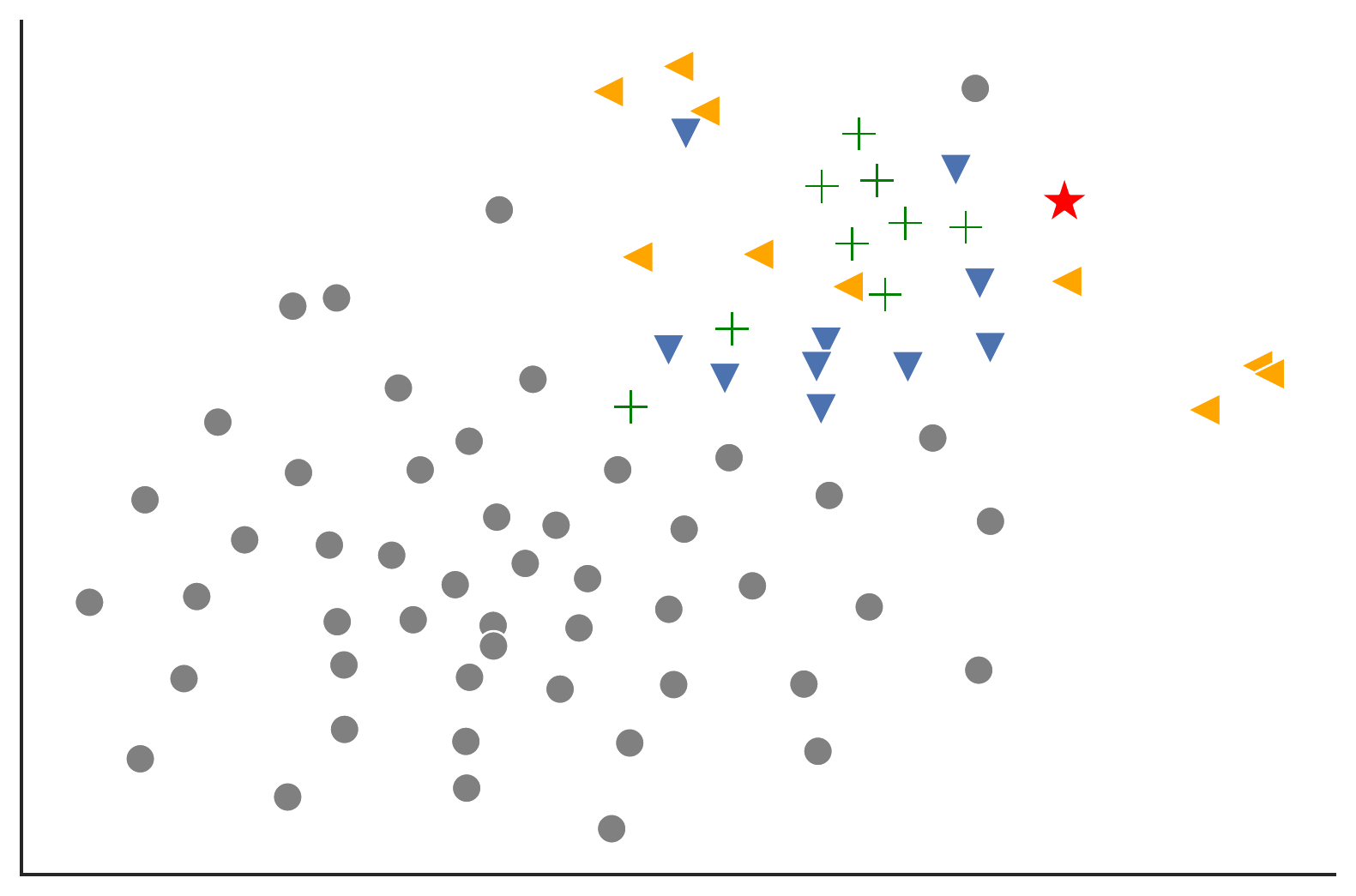}}
        \subfigure[1106007: define visceral ]{\label{case_Visceral}\includegraphics[scale=0.285]{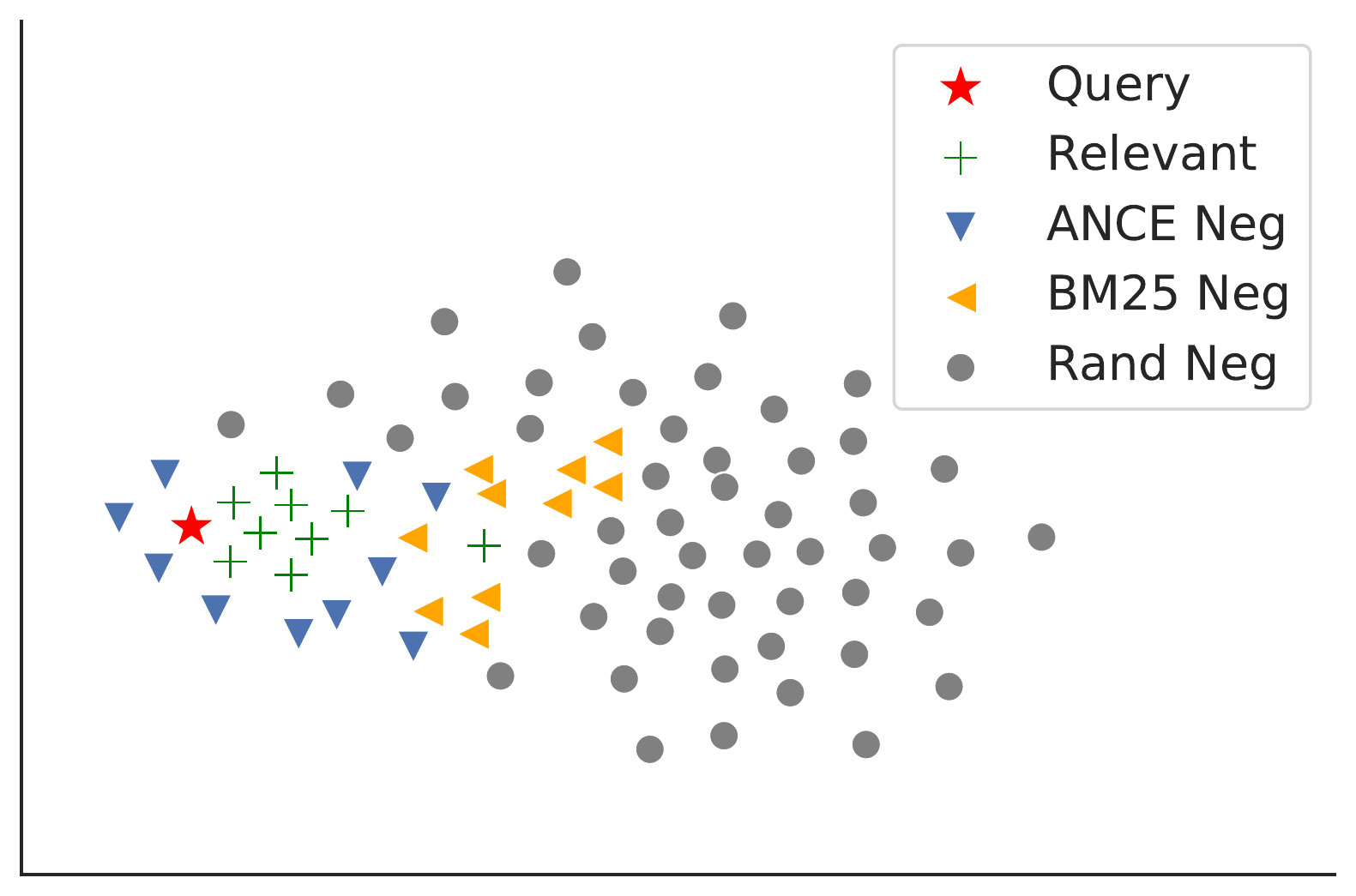}}
        \caption{t-SNE Plots for Winning Cases in Table \ref{tab:win_case_study}.
        \label{fig:t-SNE_win_case}
        }
\end{figure}

\begin{table}[t]
\centering
\small
\begin{tabular}{l|p{5cm}|p{5cm}}

 \hline\hline
   & \textbf{ANCE} 
    & \textbf{BM25}
    \\
     \hline
     
\textbf{Query:} &\multicolumn{2}{l} {qid (182539): Example of monotonic function}
\\ \hline
{Title:} &  Wikipedia: Monotonic function & Explain Extended: Things SQL needs: sargability of monotonic functions \\ \hline
{DocNo:} &D510209   &D175960   \\ \hline
{Snippet:} & In mathematics, a monotonic function (or monotone function) is a function between ordered sets that preserves or reverses the given order... For example, if y=g(x) is strictly monotonic on the range [a,b] … 
& 
I'm going to write a series of articles about the things SQL needs to work faster and more efficienly…  \\  \hline

{Ranking Position:} & 1 & 1 \\ \hline
{TREC Label:} & 0 (Irrelevant) & 2 (Relevant) \\ \hline
{NDCG@10:} & 0.25 & 0.61\\ \hline\hline

\textbf{Query:} &\multicolumn{2}{l} {qid (1117099): What is a active margin}
\\ \hline
{Title:} &  Wikipedia: Margin (finance) & Yahoo Answer: What is the difference between passive and active continental margins \\ \hline
{DocNo:} &D166625    &D2907204    \\ \hline
{Snippet:} & In finance, margin is collateral that the holder of a financial instrument … 
& 
An active continental margin is found on the leading edge of the continent where …  \\  \hline

{Ranking Position:} & 2 & 2 \\ \hline
{TREC Label:} & 0 (Irrelevant) & 3 (Very Relevant) \\ \hline
{NDCG@10:} & 0.44 & 0.74\\ \hline\hline

\textbf{Query:} &\multicolumn{2}{l} {qid (1132213): How long to hold bow in yoga}
\\ \hline
{Title:} &  Yahoo Answer: How long should you hold a yoga pose for & yogaoutlet.com: How to do bow pose in yoga \\ \hline
{DocNo:} &D3043610     &D3378723   \\ \hline
{Snippet:} & so i've been doing yoga for a few weeks now and already notice that my flexiablity has increased drastically. …That depends on the posture itself … 
& 
Bow Pose is an intermediate yoga backbend that deeply opens the chest and the front of the body…Hold for up to 30 seconds …  \\  \hline

{Ranking Position:} & 3 & 3 \\ \hline
{TREC Label:} & 0 (Irrelevant) & 3 (Very Relevant) \\ \hline
{NDCG@10:} & 0.66 & 0.74\\ \hline\hline

\end{tabular}
\caption{Queries in the TREC 2019 DL Track Document Ranking Tasks where ANCE performs worse than BM25. Snippets are manually extracted. The documents in the first position where BM25 wins are shown. The NDCG@10 of ANCE and BM25 in the corresponding query is listed. Typos in the query are from the real web search queries in TREC.}
\label{tab:loss_case_study}
\end{table}

\begin{figure}[t]
    \centering
        \subfigure[182539: monotonic function
        ]{\label{case_MonotonicFunc}\includegraphics[scale=0.285]{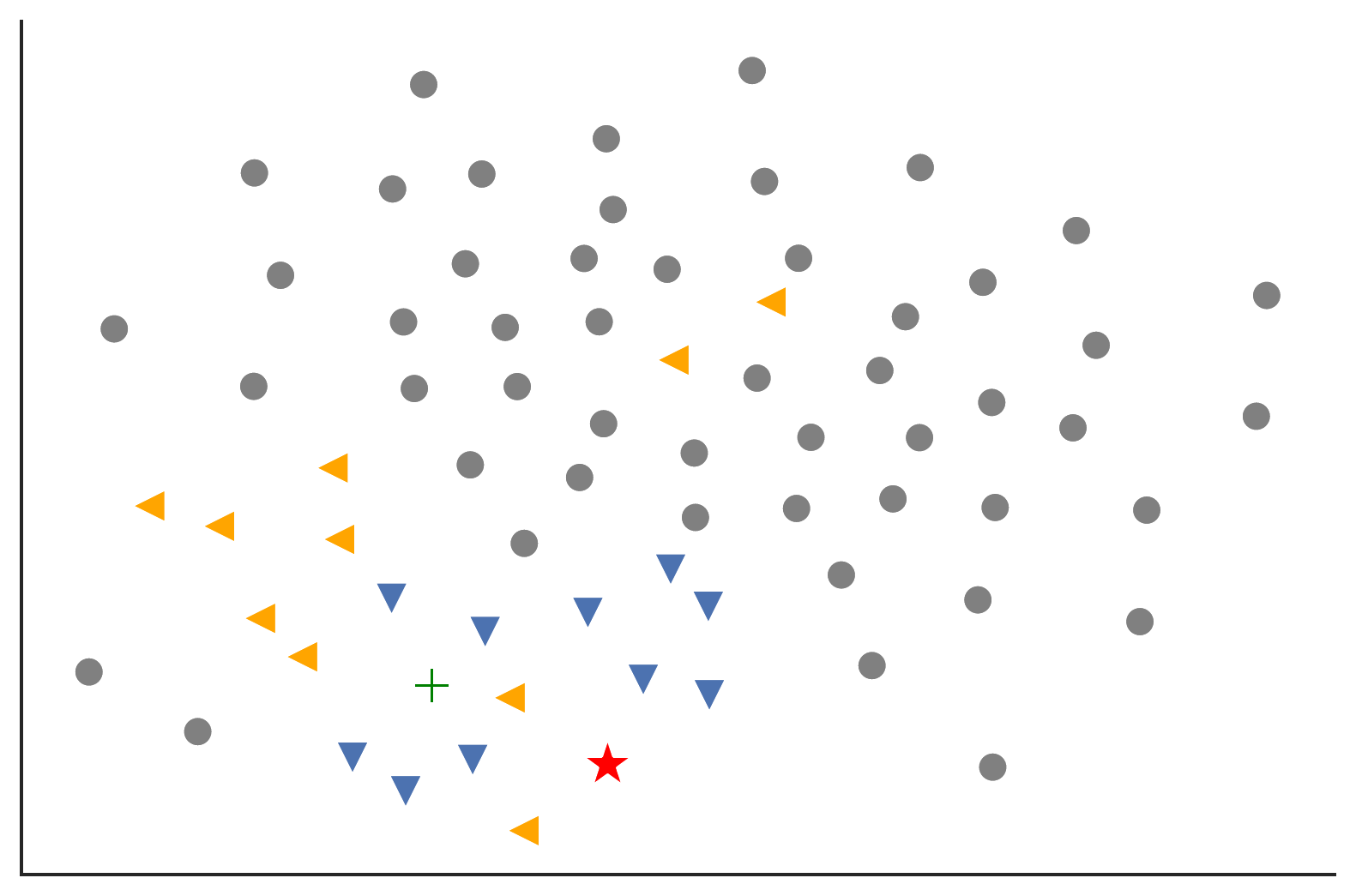}}
        \subfigure[1117099: active margin
        ]{\label{case_ActiveMargin}\includegraphics[scale=0.285]{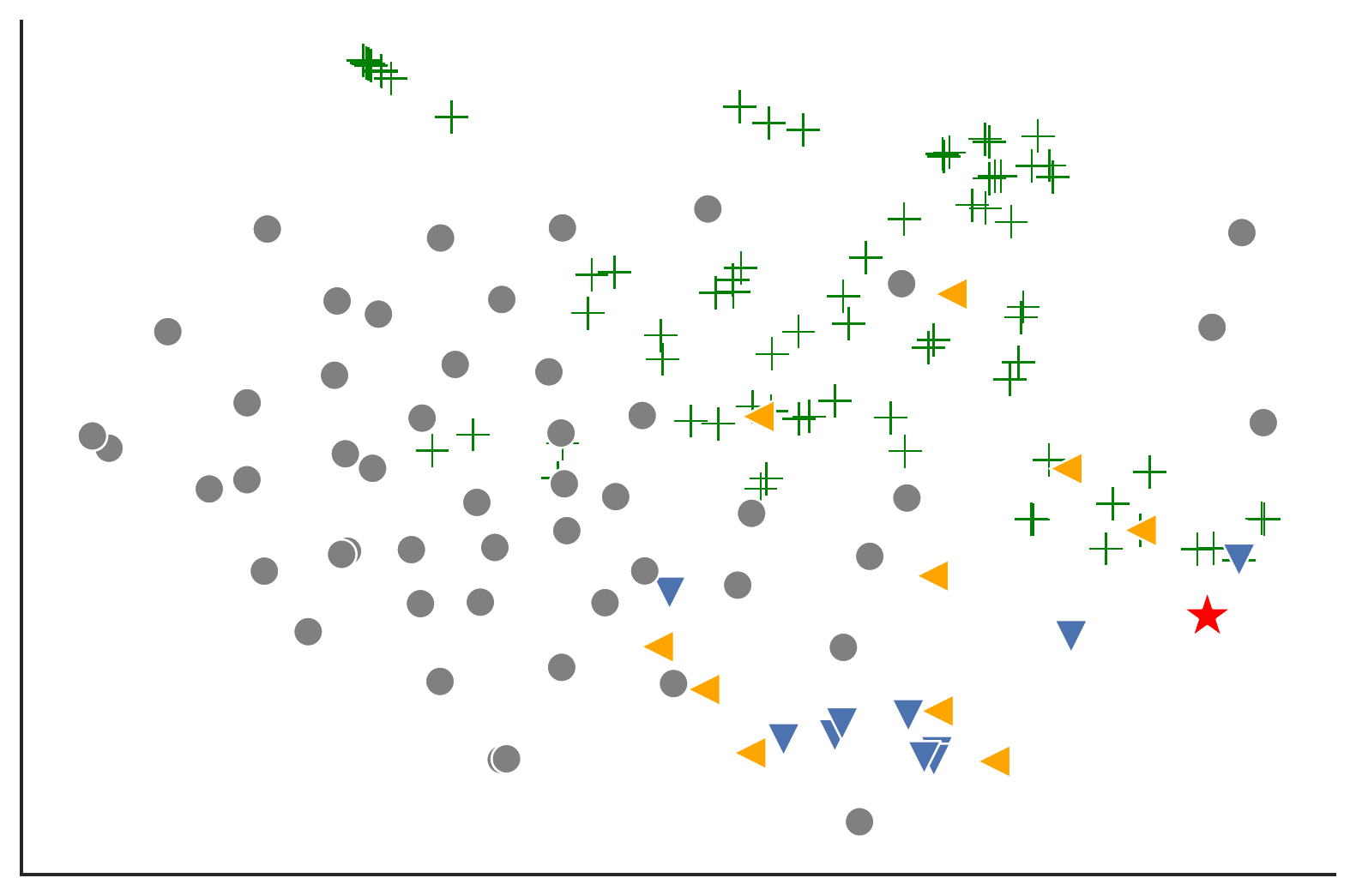}}
        \subfigure[1132213: yoga bow ]{\label{case_yoga}\includegraphics[scale=0.285]{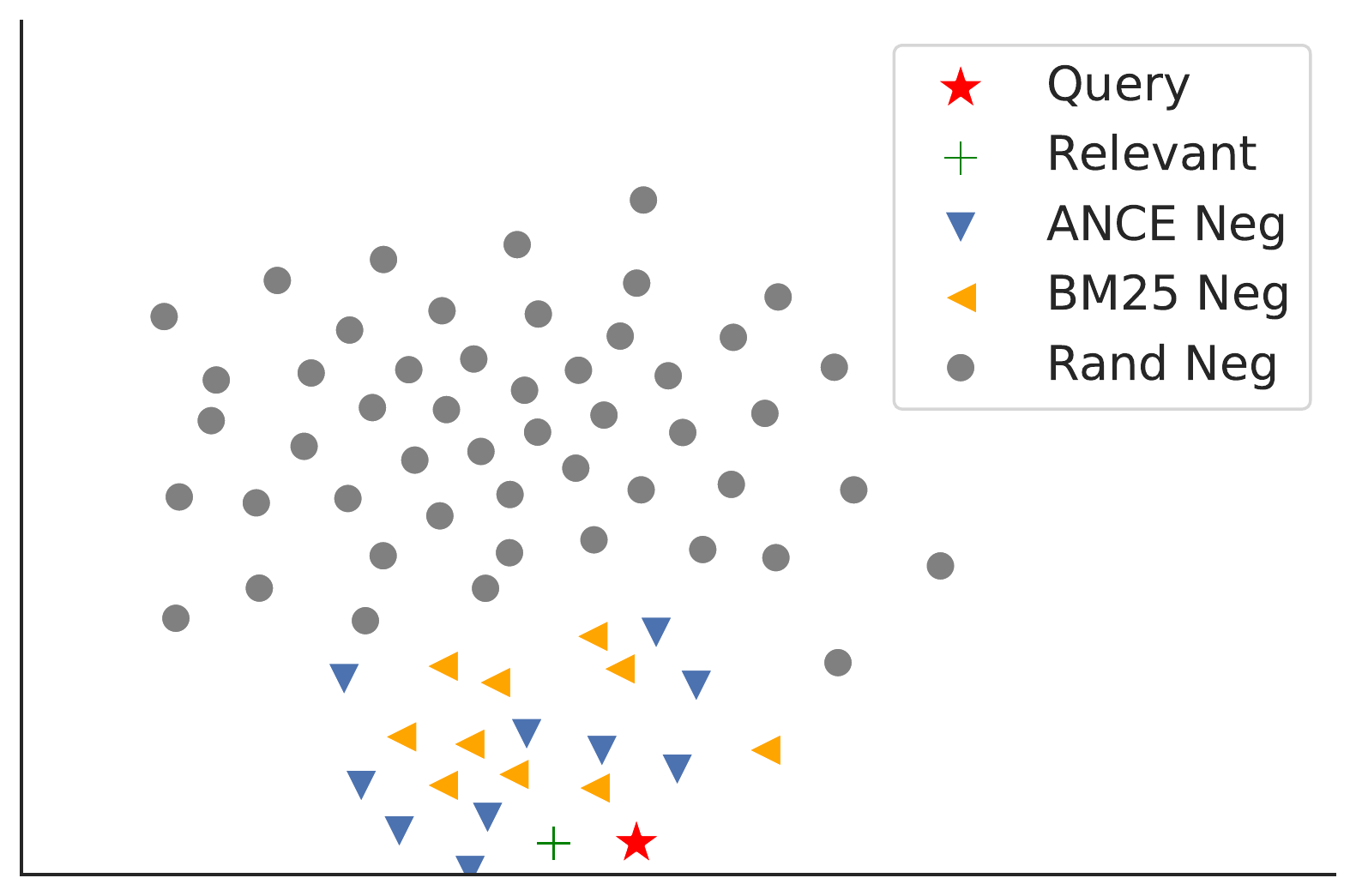}}
        \caption{t-SNE Plots for Losing Cases in Table \ref{tab:loss_case_study}.
        \label{fig:t-SNE_loss_case}
        }
\end{figure}

\subsection{Case Studies}
\label{app:case}

In this section, we show Win/Loss case studies between ANCE and BM25. 
Among the 43 TREC 2019 DL Track evaluation queries in the document task, ANCE outperforms BM25 on 29 queries, loses on 13 queries, and ties on the rest 1 query.
The winning examples are shown in Table~\ref{tab:win_case_study} and the losing ones are in Table~\ref{tab:loss_case_study}. Their corresponding ANCE-learned (FirstP) representations are illustrated by t-SNE in Fig.~\ref{fig:t-SNE_win_case} and Fig.~\ref{fig:t-SNE_loss_case}.

In general, we found ANCE better captures the semantics in the documents and their relevance to the query. The winning cases show the intrinsic limitations of sparse retrieval. For example, BM25 exact matches the ``most popular food'' in the query ``what is the most popular food in Switzerland'' but using the document is about Mexico. The term ``Switzerland'' only appears in the related question section of the web page.

The losing cases in Table~\ref{tab:loss_case_study} are also quite interesting. Many times we found that it is not that DR fails completely and retrieves documents not related to the query's information needs at all, which was a big concern when we started research in DR. The errors ANCE made include retrieving documents that are related just not exactly relevant to the query, for example, ``yoga pose'' for ``bow in yoga''. In other cases, ANCE retrieved wrong documents due to the lack of the domain knowledge: the pretrained language model may not know ``active margin'' is a geographical terminology, not a financial one (which we did not know ourselves and took some time to figure out when conducting this case study). There are also some cases where the dense retrieved documents make sense to us but were labeled irrelevant.

The t-SNE plots in Fig.~\ref{fig:t-SNE_win_case} and Fig.~\ref{fig:t-SNE_loss_case} show many interesting patterns of the learned representation space. The ANCE winning cases often correspond to clear separations of different document groups. For losing cases the representation space is more mixed, or there is too few relevant documents which may cause the variances in model performances. There are also many different interesting patterns in the ANCE-learned representation space. 
We include the t-SNE plots for all 43 TREC DL Track queries in the supplementary material.
More future analyses of the learned patterns in the representation space may help provide more insights on dense retrieval.






\end{document}